\documentclass[bibyear]{aa} 

\usepackage{xcolor}
\usepackage{graphicx}
\usepackage{txfonts}
\usepackage{booktabs}
\usepackage{siunitx}
\usepackage[colorlinks=true,citecolor=blue]{hyperref}
\usepackage{longtable}
\usepackage{amsmath}
\usepackage{upgreek}
\usepackage{multicol}
\usepackage{natbib}
\usepackage{txfonts}
\begin{document} 

\title{The expansion of the GRB\,221009A afterglow}

\author{S.~Giarratana \inst{1,2}\fnmsep\thanks{email: stefano.giarratana@inaf.it} \and O.~S.~Salafia\inst{3,4} \and M.~Giroletti\inst{2} \and G.~Ghirlanda\inst{3,4} \and L.~Rhodes\inst{5} \and P.~Atri\inst{6} \and B.~Marcote\inst{7} \and J.~Yang\inst{8} \and T.~An\inst{9,10} \and G.~Anderson\inst{11} \and J.~S.~Bright\inst{5} \and W.~Farah\inst{12} \and R.~Fender\inst{5,13} \and J.~K.~Leung\inst{14,15,16} \and S.~E.~Motta\inst{3,5} \and M.~P\'erez-Torres\inst{17,18,19} \and A.~J.~van der Horst\inst{20}}
 
    \institute{Dipartimento di Fisica e Astronomia, Università degli Studi di Bologna, Via Gobetti 93/2, 40129 Bologna, Italy    \and
    INAF Istituto di Radioastronomia, via Gobetti 101, 40129 Bologna, Italy
    \and
    INAF Osservatorio Astronomico di Brera, Via E. Bianchi 46, I-23807 Merate, Italy
    \and
    INFN -- sezione di Milano-Bicocca, Piazza della Scienza 3, I-20126 Milano (MI), Italy
    \and
    Astrophysics, Department of Physics, University of Oxford, Keble Road, Oxford OX1 3RH, UK
    \and
    ASTRON, Netherlands Institute for Radio Astronomy, Oude Hoogeveensedĳk 4, 7991 PD Dwingeloo, The Netherlands
    \and
    Joint Institute for VLBI ERIC, Oude Hoogeveensedijk 4, 7991~PD Dwingeloo, The Netherlands
    \and
    Department of Space, Earth and Environment, Chalmers University of Technology, Onsala Space Observatory, SE-439 92 Onsala, Sweden
    \and
    Shanghai Astronomical Observatory, Chinese Academy of Sciences, Nandan Road 80, Shanghai 200030, China
    \and
    Key Laboratory of Radio Astronomy and Technology, Chinese Academy of Sciences, A20 Datun Road, Beijing, 100101, P. R. China
    \and
    International Centre for Radio Astronomy Research, Curtin University, GPO Box U1987, Perth, WA 6845, Australia
    \and
    SETI Institute, 339 Bernardo Ave, Suite 200 Mountain View, CA 94043, USA
    \and
    Department of Astronomy, University of Cape Town, Private Bag X3, Rondebosch 7701, South Africa
    \and
    David A. Dunlap Department of Astronomy and Astrophysics, University of Toronto, 50 St. George Street, Toronto, Ontario, M5S 3H4, Canada
    \and
    Dunlap Institute fro Astronomy and Astrophysics, University of Toronto, 50 St. George Street, Toronto, Ontario, M5S 3H4, Canada
    \and
    Racah Institute of Physics, The Hebrew University of Jerusalem, Jerusalem, 91904, Israel
    \and
    Instituto de Astrof\'isica de Andaluc\'ia (IAA-CSIC), Glorieta de la Astronom\'ia s/n, E-18008 Granada, Spain
    \and
    Facultad de Ciencias, Universidad de Zaragoza, Pedro Cerbuna 12, E-50009 Zaragoza, Spain
    \and
    School of Sciences, European University Cyprus, Diogenes street, Engomi, 1516 Nicosia, Cyprus
    \and
    Department of Physics, George Washington University, 725 21st St NW, Washington, DC, 20052, USA
}

\date{Received \dots; accepted \dots}

 
\abstract
{
We observed $\upgamma$-ray burst (GRB) 221009A using very long baseline interferomety (VLBI) with the European VLBI Network (EVN) and the Very Long Baseline Array (VLBA), over a period spanning from 40 to 262 days after the initial GRB. The high angular resolution (mas) of our observations allowed us, for the second time ever, after GRB\,030329, to measure the projected size, $s$, of the relativistic shock caused by the expansion of the GRB ejecta into the surrounding medium. Our observations support the expansion of the shock with a $>4\sigma$-equivalent significance, and confirm its relativistic nature by revealing an apparently superluminal expansion rate. Fitting a power law expansion model, $s\propto t^a$, to the observed size evolution, we find a slope $a=0.69^{+0.13}_{-0.14}$. 
Fitting the data at each frequency separately, we find different expansion rates, pointing to a frequency-dependent behaviour. We show that the observed size evolution can be reconciled with a reverse shock plus forward shock, provided that the two shocks dominate the emission at different frequencies and, possibly, at different times.}

\keywords{Gamma-ray burst: general -- Gamma-ray burst: individual: GRB 221009A -- Radio continuum: general --  Techniques: high angular resolution -- Techniques: interferometric}

\maketitle
%

\section{Introduction}
\label{sec:intro}

On the 9 October 2022, all satellites equipped for transient detection were triggered by the extraordinary $\upgamma$-ray burst (GRB) 221009A \citep{veres2022,bissaldi2022, ursi2022, piano2022, gotz2022, frederiks2022, tan2022, mitchell2022, liu2022, lapshov2022, xiao2022, ripa2022, dichiara2022, kennea2022}. At a redshift of $z=0.151$ \citep{deugartepostigo2022, malesani2023}, GRB\,221009A holds the record of the highest ever measured isotropic equivalent energy ($E_{\gamma,\mathrm{iso}}\gtrsim 10^{55}$ erg -- \citealt{lesage2023}). 
It is the brightest GRB in the last 50\,years and it is estimated to be a one in $\sim$10\,000 years occurrence based on the observed flux distribution of other known long GRBs \citep{oconnor2023, burns2023,malesani2023}. 
Such a unique event initiated an unprecedented follow-up campaign, characterised by extensive temporal and spectral coverage. At the highest energies, the LHAASO Collaboration reported the detection of sustained emission well above 1\,TeV \citep{lhaaso2023, cao2023}. 
At the lower end of the electromagnetic spectrum, radio observations of GRB\,221009A commenced just three hours post-burst and detected the brightest ever radio counterpart, reaching a flux density of 60\,mJy \citep{bright2023}. 
Initial attempts to model the multi-wavelength afterglow emission considered contributions from both the reverse shock (RS) and the forward shock (FS) resulting from the deceleration of the ultra-relativistic jet by the surrounding material \citep{ren2023, sato2023, laskar2023, oconnor2023, gill2023, Zheng2024}. However, uncertainties persist in the final interpretation of the data, despite incorporating most of the presently known physical ingredients governing the dynamics and emission of GRB jets.

Unique measurements able to independently constrain the afterglow evolution can be obtained with milliarscsecond resolution observations. Very long baseline interferometry (VLBI) allows for direct measurements of the size of the emission region, together with high-precision astrometry. As a result, proper motion and source expansion can be measured \citep{taylor2004,mooley2018,ghirlanda2019}. If the viewing angle $\theta_\mathrm{v}$ between the observer line of sight  and the GRB jet axis is smaller than the jet half-opening angle $\theta_\mathrm{j}$ (`on-axis' GRB), the projected image during the afterglow is expected to expand, but not to show appreciable proper motion. Conversely, if the outflow is observed `off-axis' ($\theta_\mathrm{v} > \theta_\mathrm{j}$), an apparent superluminal motion is expected. 
To date, measurements of the size and expansion of the emitting region have only been possible for GRB\,030329 \citep{taylor2004,Taylor2005}, providing the first direct evidence of the relativistic expansion of GRB outflows. Over the last two decades, numerous campaigns were aimed to repeat the success of GRB\,030329 \citep[e.g.][]{nappo2017, Salafia2022_29A, Giarratana2022}. However, no event shone brightly and long enough to allow for an expansion measurement.
On the other hand, for the multi-messenger event GW\,170817 \citep{abbott2017a, abbott2017, margutti2021}, VLBI observations were fundamental to measure the apparent superluminal motion and to constrain the size of the emitting region of the non-thermal electromagnetic counterpart \citep{mooley2018,ghirlanda2019}, proving, for the first time ever, that the mergers of two neutron stars are able to successfully launch ultra-relativistic jets. 

Here we present our VLBI follow-up campaign on GRB\,221009A. The data reduction is detailed in Sec. \ref{sec:observations}. The method implemented to measure the source properties from radio observations is described in Sec. \ref{sec:methods}. In Sec. \ref{sec:results} we present the results of our campaign and discuss the physical implications in Sec. \ref{sec:discussion}. Throughout the work, we assume \citet{Planck2020} cosmological parameters. With these parameters, the angular diameter distance at $z = 0.151$ is $d_\mathrm{A}=560.3$ Mpc. Therefore, 1\,mas separation corresponds to 2.72\,pc in projection at such a distance. 

\section{VLBI observations and data reduction}
\label{sec:observations}
\subsection{European VLBI Network}
We observed the field of GRB\,221009A with the European VLBI Network (EVN) from 40 to 261 days post-burst (PI: Giarratana, project code: RG013). Given the target-of-opportunity nature of the proposal, not all antennas were available at all epochs. Table \ref{tab:antennas} lists the antennas joining each epoch. Table\,\ref{tab:log} presents a summary of the properties of the observations. The observations were performed in two different bands centred at 4.9 and 8.3\,GHz. The data were recorded at 4 Gbits s$^{-1}$. Dual polarisation products (RR, LL) were correlated at the Joint Institute for VLBI in Europe (JIVE, Dwingeloo, Netherlands) using the Super FX Correlator \citep[SFXC;][]{keimpema2015} into sixteen sub-bands with 32 MHz bandwidth and 64 channels each. 
For the last epoch, RG013 F, the data were correlated into eight sub-bands with 32 MHz bandwidth and 64 channels each. The first, EVN epoch (RG013 A), carried out 6 days post burst at a central frequency of 22.2 GHz, was not usable due to unfavourable observing conditions.

The observations consisted of phase-referencing cycles with 4.5 and 2.5 minutes on the target at 4.9 and 8.3\,GHz, respectively, and 1.5 minutes on the phase calibrator. Further scans every approximately 30 minutes on some `check' sources were also included. Throughout the observations, some scans on a fringe finder were performed. The radio source J190536.4$+$194308 (J1905$+$1943 hereafter) and the Very Large Array Sky Survey (VLASS) compact radio source J191142.50$+$195200 (J191142$+$1952 hereafter) were used as phase calibrators in the first two (RG013 B and C) and in the last three observations (RG013 D, E and F), respectively. 

The calibration was performed using {\tt AIPS}\footnote{The Astronomical Image Processing System (AIPS) is a software package produced and maintained by the National Radio Astronomy Observatory (NRAO).} \citep{Greisen2003}, following the standard procedure for EVN phase-referenced observations\footnote{\url{https://www.evlbi.org/evn-data-reduction-guide}}. The amplitude calibration, which accounts for the bandpass response, the antenna gain curves and the system temperatures, was performed by applying the gains derived by the EVN pipeline. 
We performed a correction for the dispersive delay using the IONEX files from the International GNSS Service (\texttt{vlbatecr} procedure in {\tt AIPS}), we calculated a manual single band delay on the fringe finder (\texttt{vlbampcl} procedure in {\tt AIPS}) and we carried out the global fringe fitting on the phase calibrator (\texttt{fring} task in {\tt AIPS}) using a model of the source derived by a concatenation (in {\tt CASA}, \citealt{McMullin2007}) and self-calibration (in {\tt Difmap}, \citealt{Shepherd1994}) of all the visibilities on the source obtained across the various epochs. Solutions were interpolated (\texttt{clcal} task in {\tt AIPS}) and applied to the phase calibrator itself, the target and some check sources (see Appendix\,\ref{appendix1}). For the last three epochs, we corrected the visibilities of J191142$+$1952 by fixing the phase centre in {\tt CASA} to the actual position of the phase calibrator, as the initial position of this phase calibrator was not constrained with a sub-mas resolution. 

Images of the sources were produced using {\tt Difmap}. For the analysis presented in this paper, we selected the image with the best signal-to-noise ratio (S/R) among the two images produced before and after the self-calibration of the phase calibrator, respectively. For RG013 C, the flux density of the GRB enabled a self-calibration in phase in {\tt AIPS} (solint=1\,min). Further information on the structure and the data reduction process can be found in Appendix\,\ref{appendix1}.

\subsection{Very Long Baseline Array}
The Very Long Baseline Array (VLBA) data were acquired between 44 and 262 days post-burst (PI: Atri, project code: BA160). The central frequency was 15.2 GHz, with a total bandwidth of 512 MHz, divided into 4 spectral windows of 128 MHz and 256 channels each, in full circular polarisation (RR, LL, RL, LR). The number of participating stations contributing useful data was 7, 8, 10 and 10 in experiments BA160 B, C, C1 and D respectively (see Table \ref{tab:antennas}). Each observation included approximately $30$-minute-long geodetic-style blocks at the beginning and at the end of the observation, used to determine troposphere modelling errors. The central part of the 
observations included scans on fringe finder bright calibrators and repetitions of a J1905$+$1943 -- J1925$+$2106 -- GRB~221009A sequence, where J1905$+$1943 and J1925$+$2106 are known VLBA calibrators, with respective durations of 30s -- 30s -- 80s.

The data were correlated at the NRAO in Socorro using the Distributed FX software correlator \citep[DiFX;][]{deller2011}. The data reduction was carried out in {\tt AIPS}, following standard procedures for continuum phase-referencing experiments. Procedures \texttt{vlbaeops}, \texttt{vlbaccor}, \texttt{vlbampcl}, \texttt{vlbabpss}, \texttt{vlbaamp} were carried out in this order for the initial bandpass and amplitude calibration. The following step consisted in the calibration of the troposphere modeling errors by running the task \texttt{fring} on the geodetic blocks, followed by \texttt{mbdly} and \texttt{delzn}.  The final phase, rate, and delay fringe-fitting was carried out separately on J1905$+$1943 and J1925+2106, yielding high S/R and well-behaved solutions for both sources.  The solutions from the closer phase calibrator, J1905$+$1943, were applied to the target field.  After preparing a model of the phase calibrator using {\tt Difmap}, a cycle of amplitude and phase solutions were determined for the calibrator itself and applied to the target to further refine the calibration. Finally, we produced single-source frequency-averaged datasets for the target, which were imaged in {\tt AIPS} with a natural weighting scheme.

Our VLBA campaign included one more epoch, BA160 A, at approximately 14 days post-burst. However, as the antennas were pointed at an incorrect position in the sky, the GRB fell outside the primary beam of the VLBA, which is approximately 3\,arcmin  at 15\,GHz. While the reduced sensitivity (approximately 25\% of nominal) still allowed for the detection of the burst, a satisfactory calibration of the complex visibilities was hampered. Therefore, we did not include this experiment in our analysis.

\section{Methods}
\label{sec:methods}



\begin{table*}[t]
\caption[]{Log table of our VLBI campaign and summary results of circular Gaussian fits to source visibilities.}
\centering
\resizebox{\textwidth}{!}{
\begin{tabular}{llcrccccccc|ccc}
\toprule
Code    &Date &Time  &$t_\mathrm{obs}-t_0$  &Array  &$\nu_\mathrm{obs}$    & $b_\mathrm{maj}^\dagger$ &$b_\mathrm{min}^\dagger$ &$b_\mathrm{p.a.}^\dagger$ &Phase Calibrator &r.m.s.$^\dagger$ & $F_\nu^\star$     &  FWHM$^{\star}$ & $\langle \beta_\mathrm{app}\rangle^\star$           \\ 
& &[hh:mm UT]  &[days]    &  &[GHz]  &[mas] &[mas] &[deg]   &   &[$\upmu$Jy/b] &  [mJy]  &  [mas]         & c      \\ 
\midrule
RG013 B &2022-11-18  &09:30 -- 13:30   &40   &EVN  & 8.1 -- 8.6 &1.4  &0.45 &11   &J1905$+$1943 &67 &                  ${1.02_{-0.05}^{+0.05}}$ & ${{0.12}^{+0.05}_{-0.07}}$ & ${{5.5}^{+2.5}_{-3.2}}$ \\
RG013 C &2022-11-21  &09:30 -- 13:30   &43   &EVN  &4.6 -- 5.1   &3.7 &0.69 &11    &J1905$+$1943 &16 &                 ${1.42_{-0.03}^{+0.03}}$ & ${<0.20}$ & ${<8.7}$ \\
BA160 B &2022-11-22, 23  &19:58 -- 00:58   &44   &VLBA  & 14.9 -- 15.3   & 1.4 & 0.40 &  $-9$   & J1905$+$1943 & 130 & ${0.80_{-0.07}^{+0.08}}$ & ${{0.27}^{+0.05}_{-0.06}}$ & ${{11.3}^{+2.1}_{-2.3}}$ \\
BA160 C &2023-01-31  &15:15 -- 20:14   &114   &VLBA  & 14.9 -- 15.3   &  1.4 & 0.58  & 3.5 & J1905$+$1943 & 66 &       ${0.27_{-0.05}^{+0.06}}$ & ${{0.45}^{+0.19}_{-0.19}}$ & ${{7.4}^{+3.1}_{-3.0}}$ \\
RG013 D &2023-02-03  &05:30 -- 11:30   &117   &EVN  & 4.6 -- 5.1   &7.1 &0.9 &7.8     &J191142$+$1952 &10 &            ${0.47_{-0.03}^{+0.03}}$ & ${{0.35}^{+0.09}_{-0.12}}$ & ${{5.6}^{+1.4}_{-1.9}}$ \\
RG013 E &2023-02-04  &05:30 -- 11:30   &118   &EVN  & 8.1 -- 8.6   &1.4 &0.59  &10    &J191142$+$1952 &21 &            ${0.59_{-0.06}^{+0.07}}$ & ${{0.39}^{+0.06}_{-0.07}}$ & ${{6.2}^{+0.9}_{-1.0}}$ \\
BA160 C1 &2023-05-02  &10:17 -- 15:16   &205   &VLBA  & 14.9 -- 15.3   & 1.5  & 0.51  & $-10$ & J1905$+$1943 & 35 &    ${0.33_{-0.07}^{+0.09}}$ & ${{1.39}^{+0.48}_{-0.42}}$ & ${{12.6}^{+4.3}_{-3.8}}$ \\
RG013 F &2023-06-27, 28  &19:30 -- 02:36   &261   &EVN  & 4.8 -- 5.1   &1.8 &1.5  &62    &J191142$+$1952 &10 &         ${0.16_{-0.01}^{+0.01}}$ & ${{0.42}^{+0.12}_{-0.17}}$ & ${{3.0}^{+0.9}_{-1.2}}$ \\
BA160 D &2023-06-28  &04:08 -- 09:19   &262   &VLBA  & 14.9 -- 15.3   &1.7   & 0.57  & $-3$ & J1905$+$1943 & 37 &      ${0.19_{-0.07}^{+0.19}}$ & ${<9.3}$ & ${<66}$ \\
\bottomrule
\end{tabular}}
\flushleft \footnotesize
$^\dagger$Beam major axis, minor axis, position angle, r.m.s. noise level with natural weights.\\
$^\star$Median and 68\% confidence interval of the flux density $F_\nu$ and full width at half maximum FWHM from fitting a circular Gaussian to the source visibilities; and of the average apparent expansion speed $\langle \beta_\mathrm{app}\rangle$, assuming zero size at $t_0$. If the lower extremum of the 68\% credible interval is $0$, we report the 95\% upper limit instead.
\label{tab:log}
\end{table*}

\subsection{Source flux density, size and average apparent expansion velocity estimate}\label{sec:circ_gauss_fits}

In order to extract information about the total flux density, size and position of the source from each of our epochs, we fitted a circular Gaussian source model to the calibrated visibility data adopting a Markov Chain Monte Carlo (MCMC) approach. 
The method, which is an extension of that adopted in \citet{Salafia2022_29A}, is described in
Appendix \ref{apx:corner_plots_circ_gauss_fits}. The projected angular diameter of the source image is proportional to the full width at half maximum (FWHM) of the fitted circular Gaussian, with a proportionality constant of order unity that depends on the detailed surface brightness profile \citep{Granot1999,Granot2005_GRB030329,taylor2004,Pihlstrom2007,Granot2008,Salafia2022_29A}. In what follows, we set the proportionality constant equal to 1, and discuss it whenever relevant. Once this diameter is measured, the average apparent expansion velocity can be calculated (assuming the size to be zero at the time $t_0$ of the explosion) as 
\begin{equation}
 \langle \beta_\mathrm{app}\rangle = \frac{(1+z)d_\mathrm{A}s}{2(t_\mathrm{obs}-t_0)c},
\end{equation}
where $s$ is the FWHM, $t_\mathrm{obs}$ is the time of the observation, and $c$ is the speed of light.

Table \ref{tab:log} summarises the result of the circular Gaussian fitting, along with the derived average apparent expansion velocity. In Appendix \ref{apx:corner_plots_circ_gauss_fits} we provide more detailed information in the form of corner plots that illustrate the posterior probability density of the flux density and source size from the circular Gaussian fitting. Figure \ref{fig:size_evol_all} additionally shows `violin plots' that illustrate a kernel density estimate of the posterior probability density of the FWHM for each epoch.

\subsection{Source size evolution model fitting}\label{sec:size_evol_model_fitting}

In order to fit a size evolution model $s_\mathrm{m}(t_\mathrm{obs},\vec\theta)$ to the observations, where $\vec\theta$ is a vector of free parameters, we adopted a Bayesian approach. By Bayes' theorem, and given the fact that the size estimates from different observations are independent, the posterior probability on $\vec\theta$ is proportional to the prior $\pi(\vec\theta)$ times the product of the likelihoods. This can be written as
\begin{equation}
    P\left(\vec\theta\,|\,\lbrace \vec d_{i}\rbrace_{i=1}^{M}\right) \propto \pi(\vec\theta) \prod_{i=1}^{M}\frac{P\left(s_\mathrm{m}(t_{\mathrm{obs},i})\,|\,\vec d_i\right)}{\pi(s)},
    \label{eq:size_evol_model_posterior}
\end{equation}
where M is the number of epochs included in the fit, $\vec d_i$ is the data (i.e.\ the visibilities) of the $i$-th epoch, $t_{\mathrm{obs},i}$ is the time of the $i$-th observation, $\pi(s)=\Theta(s)$ (where $\Theta$ is the Heaviside step function) is the prior on the size adopted in the circular Gaussian fits, $P(s\,|\,\vec d_i)$ is the posterior from such fits  (Eq.~\ref{eq:circ_gauss_posterior}) marginalised on all parameters except $s$. In order to evaluate Eq.~\ref{eq:size_evol_model_posterior}, we approximated the marginalised posterior on the size $P(s\,|\,\vec d_i)$ with a Gaussian kernel density estimate based on the posterior samples derived from the MCMC described in Sect.\ \ref{sec:circ_gauss_fits}. This allowed us to sample the posterior on $\vec\theta$ again through an MCMC approach. 

\subsection{Forward and reverse shock size evolution and proper motion model}\label{sec:model}

In order to interpret our observations in the context of the standard afterglow scenario, we derived a simple physical model of the size evolution and, in the case of a jet not aligned with the observer's line of sight
, the proper motion of the source expected if the emission is dominated by either the FS or the RS produced as a relativistic jet expands into an external medium with a power law number density profile $n(R) = A (R/R_\star)^{-k}$, where $R$ is the distance from the explosion site (i.e.\ the progenitor vestige) and $R_\star=5.5\times 10^{17}\,\mathrm{cm}$ is a reference radius\footnote{With this definition, $A$ has the same meaning as the usual $A_\star = 1 \,(\dot M_\mathrm{w}/10^{-5}\,\mathrm{M_\odot\,yr^{-1}})(v_\mathrm{w}/1000\,\mathrm{km\,s^{-1}})\,\mathrm{cm^{-3}}$ parameter in the wind-like external medium case ($k=2$), where $\dot M_\mathrm{w}$ and $v_\mathrm{w}$ are the mass loss rate and the velocity of the progenitor wind, assumed constant \citep{Panaitescu2000}. In the $k=0$ case, it is simply equal to the homogeneous external number density, $A=n$.}. We assumed a uniform jet angular energy profile for simplicity, with an isotropic-equivalent kinetic energy $E$, a half-opening angle $\theta_\mathrm{j}$, an initial Lorentz factor $\Gamma_0$ and a duration $T$ (which sets the jet radial width $\Delta R\sim c T$). The viewing angle is assumed to be either $\theta_\mathrm{v}=0$ (on-axis, for the calculation of the projected size) or $\theta_\mathrm{v}>\theta_\mathrm{j}$ (off-axis, for the calculation of the apparent proper motion). The model is based on the standard relativistic-hydrodynamical theory of a relativistic shock that arises from a relativistic explosion into a static, cold external medium  \citep[e.g.][]{meszaros1993,Piran1993,Sari1995,Kobayashi1999,Kobayashi2003,Yi2013} and is described in detail in Appendix \ref{apx:physical_model}. We note that the model does not include the possible sideways expansion of the shock.

The free parameters of the model are the energy-to-density ratio $E/A$, the duration $T$, the initial Lorentz factor $\Gamma_0$, the jet half-opening angle $\theta_\mathrm{j}$, the external medium density profile slope $k$ and the viewing angle $\theta_\mathrm{v}$. 
Hereafter we fix $T= T_{90}/(1+z)= 251\,\mathrm{s}$, where $T_{90}$ refers to the time encompassing the 5\% to 95\% percentile of the total photon counts as seen in the observer's reference frame, and we assume $\Gamma_0=10^{3}$ based on the lower limits from \citet{lesage2023}. Moreover, we consider only two values of the external medium density profile slope, which are $k=0$ (homogeneous external medium) and $k=2$ (wind-like external medium).

The model predicts the time evolution of the projected angular diameter of the forward and reverse shock images (Eq.\ \ref{eq:model_source_diameter}). On the other hand, our estimated source sizes are obtained by fitting a circular Gaussian model to the visibility data. The ratio of the two sizes $\xi = s_\mathrm{m}/s$ (where  $s_\mathrm{m}$ is the angular diameter predicted by the model, and $s$ is the FWHM of the circular Gaussian) depends on the lowest-order terms of the MacLaurin expansion in UV radius of the Fourier transform of the surface brightness distribution of the source \citep{Thompson2017}. Taking the surface brightness from \cite{Blandford1976} as computed in \cite{Granot2008} as reference, we expect $1.2\lesssim \xi \lesssim 1.8$. This range accommodates values previously considered in the literature, such as the value $\xi=1.4$ used in \citealt{Pihlstrom2007} and $\xi=1.3$ used in \citealt{Salafia2022_29A}. Since our model does not predict the surface brightness distribution, we include $\xi$ in our model as a nuisance parameter, with a uniform prior in the range $[1.2,1.8]$.

\section{Results}
\label{sec:results}

\subsection{Source size expansion}
\label{sec:size_expansion}

\begin{figure*}
    \centering
    \includegraphics[width=0.8\textwidth]{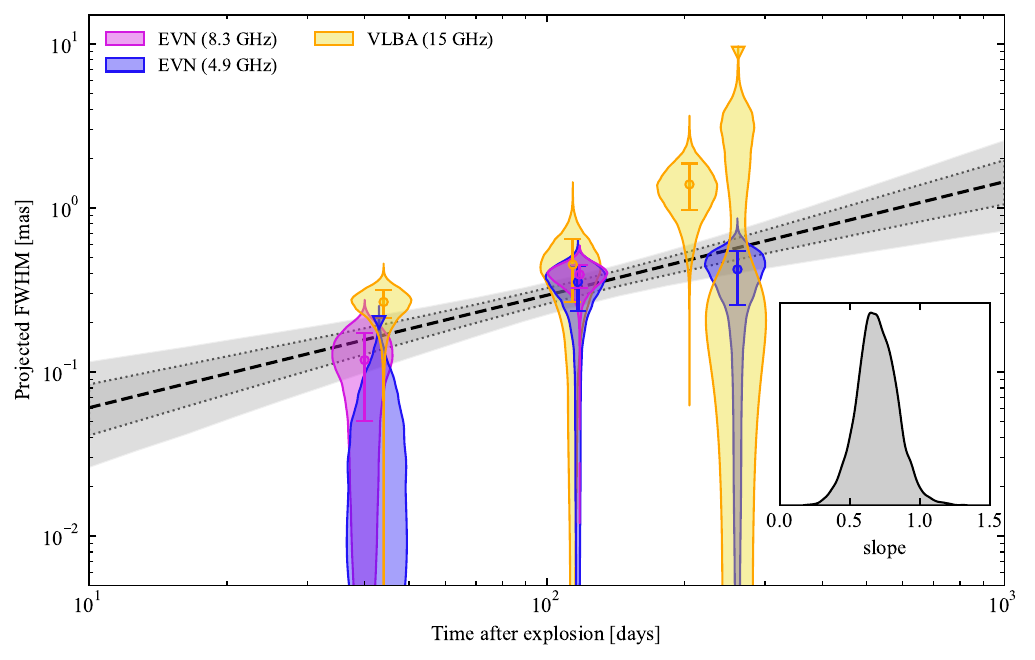}
    \caption{Source size as a function of time. The source size constraints obtained as described in Section \ref{sec:circ_gauss_fits} are shown in the form of violin plots of different colours, centred at the observing time of each epoch and of proportional width to the posterior probability density of the FWHM. In addition, we show the median and 68\% credible interval with an error bar of the same color, or the 95\% credible upper limit with a triangle if the former interval extends to 0. The black dashed line and the two grey shaded areas show respectively the median, 68\% credible interval and 95\% credible interval of the posterior predictive distribution of the source size evolution obtained from fitting a power law model $s\propto t_\mathrm{obs}^a$ to the sizes from all the epochs. The inset shows the posterior probability density of the slope $a$ from such fit.}
    \label{fig:size_evol_all}
\end{figure*}

\begin{figure}
    \centering
    
\includegraphics[width=\columnwidth]{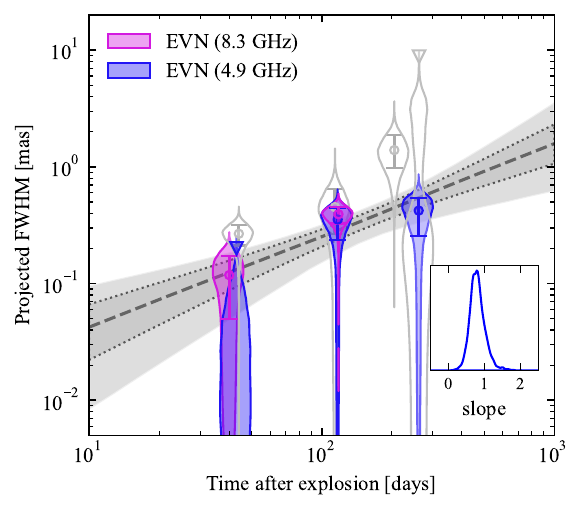}\\
    \includegraphics[width=\columnwidth]{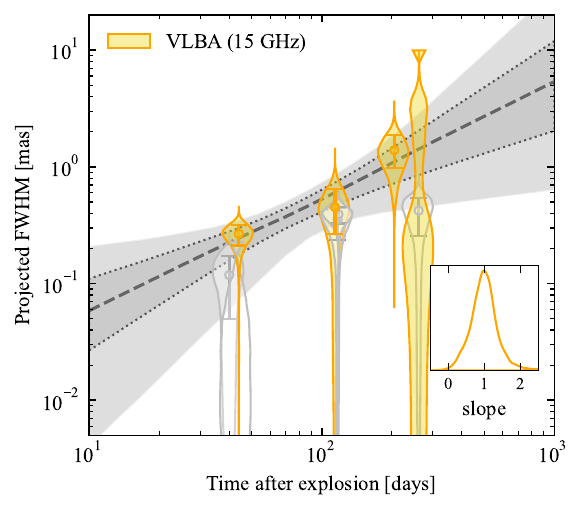}
    \caption{Size evolution considering only observations from a single array (upper panel: EVN; lower panel: VLBA). Each panel is similar to Figure \ref{fig:size_evol_all}, except that the epochs not considered in the fit are shown with light grey shading for clarity.}
    \label{fig:size_evolution_by_freq}
\end{figure}

\begin{figure}
    \centering
    \includegraphics[width=\columnwidth]{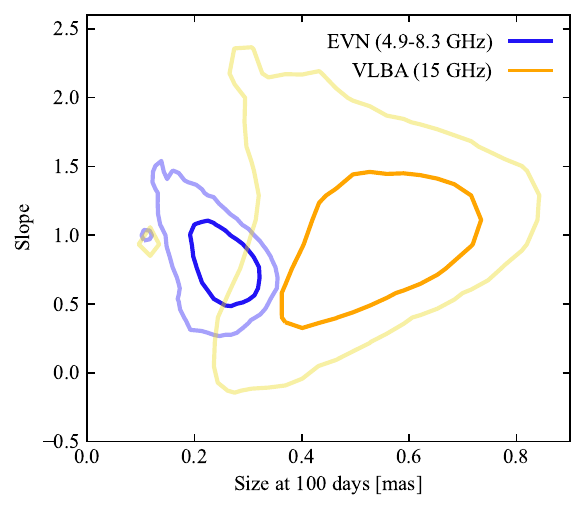}
    \caption{Comparison of the slope and normalisation of the size evolution as probed by the EVN and the VLBA. The contours in the plot contain 68\% (darker contours) and 95\% (lighter contours) of the posterior probability on the two parameters (slope and size at a reference time of 100 days) of a single power law fitted to the EVN (blue) or VLBA (orange) size evolution.}
    \label{fig:EVN_vs_VLBA_size_slope}
\end{figure}

Figure \ref{fig:size_evol_all} shows the source size constraints from Table \ref{tab:log} in the form of a `violin plot', with the width of each shaded region being proportional to the posterior probability density of the FWHM, horizontally centred at the time of the observation. Additionally, we show the median and 68\% symmetric credible interval on the FWHM by means of an error bar for each observation, except for cases where the posterior probability density does not show a clear peak, for which we show instead the $3\sigma$ upper limit with a downward-pointing triangle.
In order to quantify the source size evolution from these observations, we fit a simple phenomenological power law evolution model, $s_\mathrm{m}(t_\mathrm{obs})\propto t_\mathrm{obs}^a$, to these size measurements, through the method outlined in Sect.\ \ref{sec:size_evol_model_fitting}. The resulting posterior probability density of the power law slope is shown in the inset of Figure \ref{fig:size_evol_all}. The median and symmetric 68\% credible interval is $a = 0.69_{-0.14}^{+0.13}$. We found that more than 99.99\% of the posterior probability (>4 $\sigma$-equivalent) is located at $a>0$. Therefore, our observations strongly support the expansion of the source. In the main panel of Figure \ref{fig:size_evol_all}, we show with a black dashed line the median of the posterior predictive distribution, that is, the probability distribution of $s_\mathrm{m}(t_\mathrm{obs})$ at each fixed $t_\mathrm{obs}$, as derived from the fit. The dotted lines encompass the 68\% symmetric credible interval of the same distribution, filled with a grey shade. A lighter grey shading shows the 95\% symmetric credible interval.  
We note that the size measurements in our 15\,GHz VLBA epochs at 44 and 205\,days are in mild tension with the EVN measurements at similar times.
To explore the possibility of a frequency-dependent size, we repeated the power law size evolution model fit considering only observations performed with the EVN or VLBA. Fig.\ \ref{fig:size_evolution_by_freq} shows the resulting size evolution as fitted to EVN observations at 4.9\,GHz and 8.3\,GHz (upper panel) or VLBA observations at 15\,GHz (lower panel). The plots are similar to Figure \ref{fig:size_evol_all}, except that the epochs not considered in the fit are shown with a light grey shading. The constraint on $a$ from these fits results in medians and symmetric credible intervals of $a=0.79_{-0.23}^{+0.19}$ (4.9-8.3\,GHz) and $a=0.98_{-0.38}^{+0.36}$ (15\,GHz), in agreement with each other. On the other hand, the normalisations of the EVN and VLBA power laws differ at the $\sim 2\sigma$ level, as can be evinced from the two-dimensional posterior probabilities shown in Fig.\ \ref{fig:EVN_vs_VLBA_size_slope}. 

In order to exclude the possibility that our results with the EVN are driven by systematic effects, we carried out a series of tests including the check source J1905$+$1943. We present the results of our tests in Appendix \ref{appendix:tests}. The results of these tests indicate that the observed evolution is not driven by systematic errors in the calibrations.

\subsection{Apparent proper motion}
\label{subsec:centroid}

\begin{figure}
    \centering
    \includegraphics[width=\columnwidth]{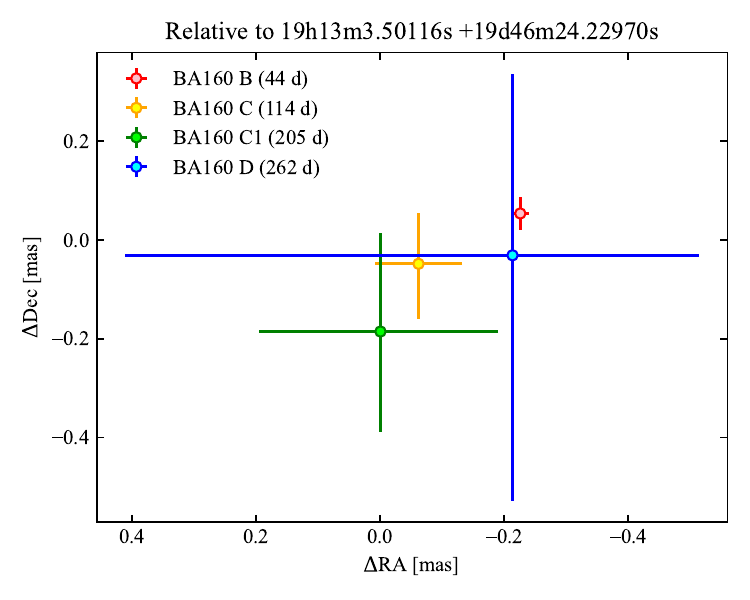}
    \caption{Source position in our VLBA observations. For each epoch, we show the statistical error bar centered on the median position from the circular Gaussian source fit, with bars spanning the symmetric 68\% credible interval on the source position in each direction.}
    \label{fig:displacement_plot}
\end{figure}

VLBI observations can constrain the apparent proper motion of the centroid of the emission and, therefore, the jet viewing angle. 
The source position at each VLBA epoch is displayed in Fig.\ \ref{fig:displacement_plot}: our results do not show any significant apparent proper motion between 44 and 262 days post-burst, but our statistical errors can accommodate a displacement of up to about 0.6 mas (at the one-$\sigma$ level) over that period. As shown in Appendix \ref{apx:proper_motion_model}, such an upper limit does not constrain strongly $\theta_\mathrm{v}$, which can still be several degrees off the edge of the jet, unless the energy-to-density ratio of the explosion is very large. Still, a number of studies including \citet{lhaaso2023} and \citet{oconnor2023} have used their data to justify a very small $\theta_\mathrm{v}$ for GRB\,221009A, indicating that we are viewing the jet close to on-axis. The lack of significant proper motion observed during our VLBI campaign is fully consistent with such on-axis scenario.

We note that the EVN campaign was not used for such study because of the change in phase reference source between the second and third epoch. While this change was motivated by the discovery of a closer phase calibrator (and hence a more efficient observing strategy), the different systematics and the lack of a reliable a priori position of the new calibrator prevent a reliable astrometric characterisation.
%
%

\section{Discussion}
\label{sec:discussion}

\subsection{Slope of the size evolution with time}

\begin{figure}
    \centering
    \includegraphics[width = \columnwidth]{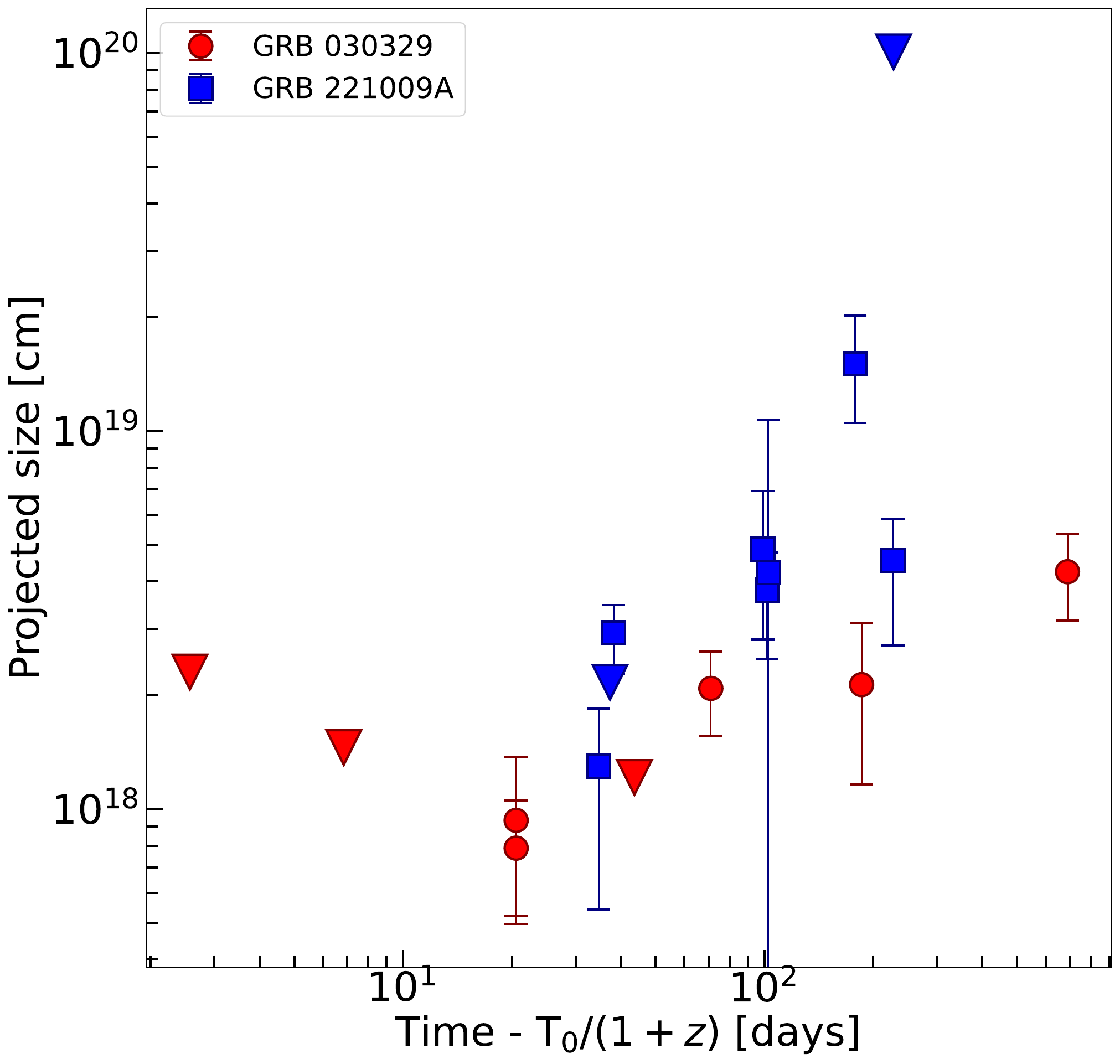}
    \caption{Comparison of size measurements from VLBI observations of GRB 030329 (red circles; \citealt{taylor2004, Pihlstrom2007}) and GRB 221009A (dark blue squares, this work). Triangles represent upper limits.}
    \label{fig:030329}
\end{figure}

The size evolution power law slope $a = 0.69_{-0.14}^{+0.13}$ we derived is 
compatible 
with the expected slopes for a spherical \citet{Blandford1976} blastwave expanding into a homogeneous medium, $\alpha=5/8=0.625$, or a wind-like medium, $\alpha=3/4=0.75$. 
Moreover, the evolution of the projected physical size is quite similar to that of 
GRB\,030329, the only other burst to date with a measured expansion rate \citep[][Fig.\ref{fig:030329}]{taylor2004}.
%
%

On the other hand, at the time of our observations, the anisotropy of the shock due to the finite opening angle of the jet should be observable, both in terms of a steepening (a `jet break', \citealt{Rhoads1997}) in the light curves, and in terms of a flattening in the evolution of the projected size \citep{Granot2005_GRB030329}. The presence of a jet break in the very high energy afterglow light curve at $<$ 1000 seconds post-trigger has been discussed by \cite{lhaaso2023}. A jet-break was also suggested by \cite{levan2023} at $\lesssim 2600$ seconds post-trigger, using optical to mid-IR data. 
The expected post-jet-break size evolution slope in the case of a homogeneous external medium is $a=1/4$, in absence of an efficient sideways expansion of the shock \citep{Granot2005_GRB030329}. Such a shallow slope is in tension with the observed one at the $\sim 3\sigma$ level. Conversely, the expected average slope is steeper and lies in the range $\langle a \rangle \sim 0.6-0.8$ if the shock expands sideways \citep{Granot2005_GRB030329}, which is compatible with the observed one. 
Therefore, in the homogeneous external medium scenario, our observations indicate that either the jet break has not happened yet, or that the shock is expanding sideways. Given the very large isotropic equivalent energy in the gamma-rays, the `late jet break' scenario would pose very demanding requirements on the total energy \citep[see e.g.][]{oconnor2023}. On the other hand, numerical simulations of external shocks arising from relativistic jets and analytical arguments seem to indicate that the sideways expansion is inefficient, unless the initial opening angle is very narrow \citep{vanEerten2010,DeColle2012,Granot2012}. These difficulties could be alleviated if the jet features a structure consisting of a narrow `core' surrounded by `wings' where the kinetic energy per unit solid angle decreases slowly, as suggested by \citet{oconnor2023} and \citet{gill2023}. This profile would steepen the evolution of the observed size, making it more similar to the spherical case, but with a reduced energy requirement with respect to a wide jet with a uniform angular energy profile.

In the wind medium case, the expected post-jet-break size evolution slope is $a=1/2$ in absence of sideways expansion. Therefore, in such a scenario, the observed evolution does not indicate the need for sideways expansion nor for a non-uniform structure within the opening angle.


\subsection{Possible frequency-dependent size}

\begin{figure}
    \centering
    \includegraphics[width=\columnwidth]{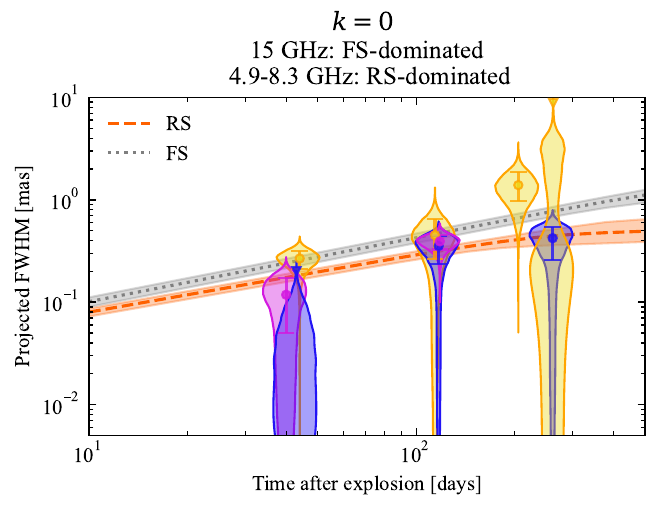}\\
    \includegraphics[width=\columnwidth]{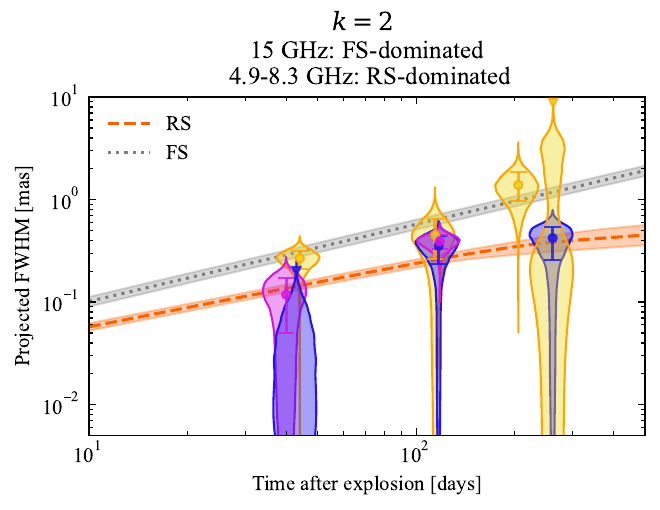}\\
    \caption{Model size evolution in a FS plus RS scenario. The violins and the error bars show the source size evolution as inferred by our observations, in the same way as in Figure \ref{fig:size_evol_all}. The gray dotted line and orange dashed lines show the medians of the posterior predictive distributions of the FS and RS size, respectively, as obtained by fitting the physical model described in Appendix \ref{apx:physical_model} to the sizes shown in the figure, assuming a homogeneous (top panel) or wind-like (bottom panel) external medium, and assuming the FS to dominate at 15\,GHz and the RS to dominate at 4.9 and 8.3\,GHz. The shaded bands around these lines show the 68\% credible interval of the posterior predictive distribution.}
    \label{fig:physical_model_RS+FS}
\end{figure}

\begin{figure}
    \centering
    \includegraphics[width=\columnwidth]{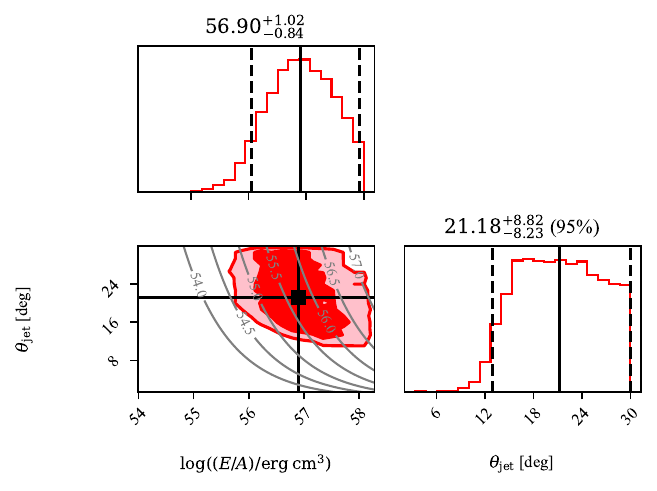}\\
    \includegraphics[width=\columnwidth]{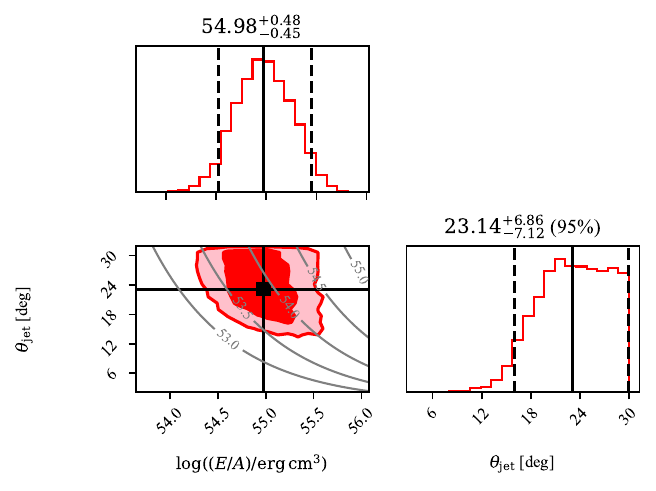}\\
    \caption{Corner plot of the posterior probability density of the physical model parameters in the homogeneous (top panel) or wind-like (bottom panel) external medium case and assuming the FS to dominate at 15\,GHz and the RS to dominate at $4.9$--$8.3$ GHz. In each panel, the red histograms show the marginalised posterior probability densities, with black solid vertical lines showing the median and dashed lines showing the 68\% credible interval or, if the latter extends to the lower (upper) extremum of the prior range, the 95\% upper (lower) limit (values reported on top of the panels). The filled contours show the smallest regions containing 68\% and 95\% of the two-dimensional posterior probability, with the black squares showing the position of the median. The grey lines show contours of constant total jet energy assuming $A=1\,\mathrm{cm^{-3}}$. Each contour is labeled with the base-10 logarithm of the corresponding total jet energy.}
    \label{fig:phys_model_corner}
\end{figure}


As discussed in section \ref{sec:size_expansion}, our data suggest the presence of a frequency-dependent size evolution. We explore here a possible avenue to interpret this behaviour. The radio afterglow of this GRB cannot be explained by a simple FS propagating either into a wind-like or a homogeneous environment \citep{ren2023, sato2023, Ren2023b, Zheng2024}. Using a data set encompassing observations from the GeV to the radio domain, \cite{laskar2023} showed that the standard afterglow model struggles at explaining the radio emission both with a FS and a RS of a conical jet propagating through a wind-like environment, leading them to invoke an additional component whose temporal evolution does not follow the standard prescriptions. Such a component could be a RS, dominating the emission at the lower frequencies ($\lesssim 10$ GHz) up to $t_\mathrm{obs}\lesssim 100\,\mathrm{d}$ \citep[see the modelling of][]{oconnor2023,gill2023}. 

Stimulated by these studies, we explored a scenario where the emission we observed is a superposition of a FS and a RS. We fitted the model described in Sect.\ \ref{sec:model} to the observed size evolution, leaving $E/A$ and $\theta_\mathrm{j}$ as our free parameters, and additionally including the $\xi$ nuisance parameter (see section \ref{sec:model}). The external medium power law index was fixed to $k=0$ or $k=2$. Based on the apparent possible frequency-dependent behaviour, we assumed the higher-frequency observations (15\,GHz) to be dominated by the FS, while the lower-frequency ones (4.9 -- 8.3\,GHz) to be dominated by the RS. For each external density profile, the posterior probability density of $(E/A,\theta_\mathrm{j},\xi)$ was derived through the Bayesian approach described in Sect.\ \ref{sec:size_evol_model_fitting}, with uniform-in-log priors on $E/A$ in the range $[10^{52},10^{58}]$ erg cm$^3$ and on $\theta_\mathrm{j}$ in the range $[0.5,30]$ deg, and a uniform prior on $\xi$ in the range $[1.2,1.8]$, as explained in section \ref{sec:model}. Figure \ref{fig:physical_model_RS+FS} shows the FS and RS model size evolution fitted to the observations. Figure \ref{fig:phys_model_corner} shows corner plots of the posterior probability densities, marginalized over $\xi$.

The model can reasonably reproduce the data in both the homogeneous ($k=0$) and wind-like ($k=2$) external medium cases. Due to the relatively steep observed size evolution, and given that our model does not include sideways expansion, the required jet half-opening angle is large ($\theta_\mathrm{j}> 13\,\mathrm{deg}$ at the 95\% credible level for $k=0$; $\theta_\mathrm{j}> 16\,\mathrm{deg}$ for $k=2$). Assuming the standard reference external density $A=1\,\mathrm{cm^{-3}}$, this in turn pushes the total jet energy $E_\mathrm{jet}=E(1-\cos\theta_\mathrm{j})$ to very large values, as demonstrated by the grey contours in Fig.\ \ref{fig:phys_model_corner}. The energy requirement can be reduced if the external density is much lower than the reference value, $A\ll 1\,\mathrm{cm^{-3}}$, or if the jet features an angular structure such that the energy per unit solid angle decreases away from the jet axis, as discussed in the previous section.  In this case, the decrease must be shallow ($E\propto \theta^{-a}$ with $a<2$, where $\theta$ is the angle from the jet axis), otherwise both the emission and the size evolution would simply reflect those of a uniform jet \citep{oconnor2023}, leading again to the same difficulties with the jet opening angle and total energy.

\section{Conclusions}
\label{sec:conclusions}
In this paper, we presented VLBI observations of the brightest $\upgamma$-ray burst ever observed, GRB\,221009A. The high angular resolution provided by the EVN and the VLBA allowed us to constrain the size and the expansion of the blast wave produced by the GRB ejecta for the second time ever. The expansion rate is consistent with the expectation for a spherical \citet{Blandford1976} blast wave (i.e., an ultra-relativistic FS) propagating into a homogeneous or wind-like medium. This could be taken as an indication that the shock is anisotropic only on angular scales larger than those probed by our observations (i.e.\ the jet break has not happened yet). This in turn points to an extremely large total (collimation-corrected) energy in the jet, especially if the external medium features a homogeneous density. The demanding energy requirement could be alleviated if the external density were much lower than usually assumed, or if the jet energy per unit solid angle decreased slowly with the angle from the jet axis, as proposed by \cite{oconnor2023} and \cite{gill2023}. Alternatively, the shock could be undergoing sideways expansion during the time covered by our observations.

Additionally, our observations suggest a frequency-dependent size evolution, with the VLBA observations at 15\,GHz showing a somewhat larger size at 40 days after the explosion, and a faster increase afterwards, with respect to the EVN observations at 5 and 8\,GHz. This could be due to the emission being dominated by the reverse shock at the lower frequencies, and by the forward shock at the higher frequencies.
Our work highlights the crucial role played by multi-wavelength VLBI monitoring of transient events both at early and late times, and in providing a vital insight into the physics of such events.

\begin{acknowledgements}

The European VLBI Network is a joint facility of independent European, African, Asian, and North American radio astronomy institutes. Scientific results from data presented in this publication are derived from the following EVN project code: RG013.

The National Radio Astronomy Observatory is a facility of the National Science Foundation operated under cooperative agreement by Associated Universities, Inc.

This work made use of the Swinburne University of Technology software correlator, developed as part of the Australian Major National Research Facilities Programme and operated under licence.

SG would like to thank Z. Paragi and the staff of JIVE for their help and support during his visiting period in Dwingeloo. We would like to thank the directors and staff of all the EVN telescopes for approving, executing, and processing our out-of-session ToO observations.

The research leading to these results has received funding from the European Union's Horizon 2020 Research and Innovation Programme under grant agreement No. 101004719 (OPTICON RadioNet Pilot).

The research leading to these results has received funding from the European Union’s Horizon 2020 Programme under the AHEAD2020 project (grant agreement n. 871158).

This work has been funded by the European Union-Next Generation EU, PRIN 2022 RFF M4C21.1 (202298J7KT - PEACE).

OS acknowledges funding from the Istituto Nazionale di Astrofisica, project number 1.05.23.04.04. 

BM acknowledges financial support from the State Agency for Research of the Spanish Ministry of Science and Innovation under grant PID2019-105510GB-C31/AEI/10.13039/501100011033 and through the Unit of Excellence Mar\'ia de Maeztu 2020--2023 award to the Institute of Cosmos Sciences (CEX2019- 000918-M).

MPT acknowledges financial support through grants CEX2021-001131-S and
PID2020-117404GB-C21 funded by the Spanish MCIN/AEI/ 10.13039/501100011033.
\end{acknowledgements}

%
%
\bibliographystyle{aa}
\bibliography{bibliography}

\newcommand{\noop}[1]{}
\begin{thebibliography}{70}
\expandafter\ifx\csname natexlab\endcsname\relax\def\natexlab#1{#1}\fi

\bibitem[{{Abbott} {et~al.}(2017{\natexlab{a}}){Abbott}, {Abbott}, {Abbott}, {Acernese}, {Ackley}, {Adams}, {Adams}, {Addesso}, {Adhikari}, {Adya}, {Affeldt}, {Afrough}, {Agarwal}, {Agathos}, {Agatsuma}, {Aggarwal}, {Aguiar}, {Aiello}, {Ain}, {Ajith}, {Allen}, {Allen}, {Allocca}, {Altin}, {Amato}, {Ananyeva}, {Anderson}, {Anderson}, {Angelova}, {Antier}, {Appert}, {Arai}, {Araya}, {Areeda}, {Arnaud}, {Arun}, {Ascenzi}, {Ashton}, {Ast}, {Aston}, {Astone}, {Atallah}, {Aufmuth}, {Aulbert}, {AultONeal}, {Austin}, {Avila-Alvarez}, {LIGO Scientific Collaboration}, \& {Virgo Collaboration}}]{abbott2017a}
{Abbott}, B.~P., {Abbott}, R., {Abbott}, T.~D., {et~al.} 2017{\natexlab{a}}, \prl, 119, 161101

\bibitem[{{Abbott} {et~al.}(2017{\natexlab{b}}){Abbott}, {Abbott}, {Abbott}, {Acernese}, {Ackley}, {Adams}, {Adams}, {Addesso}, {Adhikari}, {Adya}, \& et~al.}]{abbott2017}
{Abbott}, B.~P., {Abbott}, R., {Abbott}, T.~D., {et~al.} 2017{\natexlab{b}}, \apjl, 848, L13

\bibitem[{{Bissaldi} {et~al.}(2022){Bissaldi}, {Omodei}, {Kerr}, \& {Fermi-LAT Team}}]{bissaldi2022}
{Bissaldi}, E., {Omodei}, N., {Kerr}, M., \& {Fermi-LAT Team}. 2022, GRB Coordinates Network, 32637, 1

\bibitem[{{Blandford} \& {McKee}(1976)}]{Blandford1976}
{Blandford}, R.~D. \& {McKee}, C.~F. 1976, Physics of Fluids, 19, 1130

\bibitem[{{Bright} {et~al.}(2023){Bright}, {Rhodes}, {Farah}, {Fender}, {van der Horst}, {Leung}, {Williams}, {Anderson}, {Atri}, {DeBoer}, {Giarratana}, {Green}, {Heywood}, {Lenc}, {Murphy}, {Pollak}, {Premnath}, {Scott}, {Sheikh}, {Siemion}, \& {Titterington}}]{bright2023}
{Bright}, J.~S., {Rhodes}, L., {Farah}, W., {et~al.} 2023, Nature Astronomy, 7, 986

\bibitem[{{Burns} {et~al.}(2023){Burns}, {Svinkin}, {Fenimore}, {Kann}, {Ag{\"u}{\'\i} Fern{\'a}ndez}, {Frederiks}, {Hamburg}, {Lesage}, {Temiraev}, {Tsvetkova}, {Bissaldi}, {Briggs}, {Dalessi}, {Dunwoody}, {Fletcher}, {Goldstein}, {Hui}, {Hristov}, {Kocevski}, {Lysenko}, {Mailyan}, {Mangan}, {McBreen}, {Racusin}, {Ridnaia}, {Roberts}, {Ulanov}, {Veres}, {Wilson-Hodge}, \& {Wood}}]{burns2023}
{Burns}, E., {Svinkin}, D., {Fenimore}, E., {et~al.} 2023, \apjl, 946, L31

\bibitem[{{Cao} {et~al.}(2023){Cao}, {Aharonian}, {An}, {Axikegu}, {Bai}, {Bao}, {Bastieri}, {Bi}, {Bi}, {Cai}, {Cao}, {Cao}, {Cao}, {Chang}, {Chang}, {Chen}, {Chen}, {Chen}, {Chen}, {Chen}, {Chen}, {Chen}, {Chen}, {Chen}, {Chen}, {Chen}, {Chen}, {Cheng}, {Cheng}, {Cui}, {Cui}, {Cui}, {Cui}, {Dai}, {Dai}, {Dai}, {Danzengluobu}, {della Volpe}, {Dong}, {Duan}, {Fan}, {Fan}, {Fang}, {Fang}, {Feng}, {Feng}, {Feng}, {Feng}, {Feng}, {Gabici}, {Gao}, {Gao}, {Gao}, {Gao}, {Gao}, {Gao}, {Ge}, {Geng}, {Giacinti}, {Gong}, {Gou}, {Gu}, {Guo}, {Guo}, {Guo}, {Guo}, {Han}, {He}, {He}, {He}, {He}, {He}, {Heller}, {Hor}, {Hou}, {Hou}, {Hou}, {Hu}, {Hu}, {Hu}, {Huang}, {Huang}, {Huang}, {Huang}, {Huang}, {Huang}, {Huang}, {Ji}, {Jia}, {Jia}, {Jiang}, {Jiang}, {Jiang}, {Jin}, {Kang}, {Ke}, {Kuleshov}, {Kurinov}, {Li}, {Li}, {Li}, {Li}, {Li}, {Li}, {Li}, {Li}, {Li}, {Li}, {Li}, {Li}, {Li}, {Li}, {Li}, {Li}, {Li}, {Li}, {Li}, {Liang}, {Liang}, {Lin}, {Liu}, {Liu}, {Liu}, {Liu}, {Liu}, {Liu}, {Liu}, {Liu}, {Liu}, {Liu}, {Liu},
  {Liu}, {Liu}, {Liu}, {Lu}, {Luo}, {Lv}, {Ma}, {Ma}, {Ma}, {Mao}, {Min}, {Mitthumsiri}, {Mu}, {Nan}, {Neronov}, {Ou}, {Pang}, {Pattarakijwanich}, {Pei}, {Qi}, {Qi}, {Qiao}, {Qin}, {Ruffolo}, {S{\'a}iz}, {Semikoz}, {Shao}, {Shao}, {Shchegolev}, {Sheng}, {Shu}, {Song}, {Stenkin}, {Stepanov}, {Su}, {Sun}, {Sun}, {Sun}, {Tam}, {Tang}, {Tang}, {Tian}, {Wang}, {Wang}, {Wang}, {Wang}, {Wang}, {Wang}, {Wang}, {Wang}, {Wang}, {Wang}, {Wang}, {Wang}, {Wang}, {Wang}, {Wang}, {Wang}, {Wang}, {Wang}, {Wang}, {Wang}, {Wang}, {Wei}, {Wei}, {Wei}, {Wen}, {Wu}, {Wu}, {Wu}, {Wu}, {Wu}, {Xi}, {Xia}, {Xia}, {Xiang}, {Xiao}, {Xiao}, {Xin}, {Xin}, {Xing}, {Xiong}, {Xu}, {Xu}, {Xu}, {Xu}, {Xue}, {Yan}, {Yan}, {Yan}, {Yang}, {Yang}, {Yang}, {Yang}, {Yang}, {Yang}, {Yang}, {Yang}, {Yang}, {Yao}, {Yao}, {Ye}, {Yin}, {Yin}, {You}, {You}, {Yu}, {Yuan}, {Yue}, {Zeng}, {Zeng}, {Zeng}, {Zha}, {Zhang}, {Zhang}, {Zhang}, {Zhang}, {Zhang}, {Zhang}, {Zhang}, {Zhang}, {Zhang}, {Zhang}, {Zhang}, {Zhang}, {Zhang}, {Zhang}, {Zhang}, {Zhang},
  {Zhang}, {Zhang}, {Zhao}, {Zhao}, {Zhao}, {Zhao}, {Zhao}, {Zheng}, {Zhou}, {Zhou}, {Zhou}, {Zhou}, {Zhou}, {Zhou}, {Zhou}, {Zhu}, {Zhu}, {Zhu}, {Zhu}, \& {Zuo}}]{cao2023}
{Cao}, Z., {Aharonian}, F., {An}, Q., {et~al.} 2023, Science Advances, 9, eadj2778

\bibitem[{{Chevalier} \& {Li}(2000)}]{chevalier2000}
{Chevalier}, R.~A. \& {Li}, Z.-Y. 2000, \apj, 536, 195

\bibitem[{{De Colle} {et~al.}(2012){De Colle}, {Granot}, {L{\'o}pez-C{\'a}mara}, \& {Ramirez-Ruiz}}]{DeColle2012}
{De Colle}, F., {Granot}, J., {L{\'o}pez-C{\'a}mara}, D., \& {Ramirez-Ruiz}, E. 2012, \apj, 746, 122

\bibitem[{{de Ugarte Postigo} {et~al.}(2022){de Ugarte Postigo}, {Izzo}, {Pugliese}, {Xu}, {Schneider}, {Fynbo}, {Tanvir}, {Malesani}, {Saccardi}, {Kann}, {Wiersema}, {Gompertz}, {Thoene}, {Levan}, \& {Stargate Collaboration}}]{deugartepostigo2022}
{de Ugarte Postigo}, A., {Izzo}, L., {Pugliese}, G., {et~al.} 2022, GRB Coordinates Network, 32648

\bibitem[{{Deller} {et~al.}(2011){Deller}, {Brisken}, {Phillips}, {Morgan}, {Alef}, {Cappallo}, {Middelberg}, {Romney}, {Rottmann}, {Tingay}, \& {Wayth}}]{deller2011}
{Deller}, A.~T., {Brisken}, W.~F., {Phillips}, C.~J., {et~al.} 2011, \pasp, 123, 275

\bibitem[{{Dichiara} {et~al.}(2022){Dichiara}, {Gropp}, {Kennea}, {Kuin}, {Lien}, {Marshall}, {Tohuvavohu}, {Williams}, \& {Neil Gehrels Swift Observatory Team}}]{dichiara2022}
{Dichiara}, S., {Gropp}, J.~D., {Kennea}, J.~A., {et~al.} 2022, GRB Coordinates Network, 32632

\bibitem[{{Foreman-Mackey} {et~al.}(2013){Foreman-Mackey}, {Hogg}, {Lang}, \& {Goodman}}]{Foreman-Mackey2013}
{Foreman-Mackey}, D., {Hogg}, D.~W., {Lang}, D., \& {Goodman}, J. 2013, \pasp, 125, 306

\bibitem[{{Frederiks} {et~al.}(2022){Frederiks}, {Lysenko}, {Ridnaia}, {Svinkin}, {Tsvetkova}, {Ulanov}, {Cline}, \& {Konus-Wind Team}}]{frederiks2022}
{Frederiks}, D., {Lysenko}, A., {Ridnaia}, A., {et~al.} 2022, GRB Coordinates Network, 32668

\bibitem[{{Ghirlanda} {et~al.}(2019){Ghirlanda}, {Salafia}, {Paragi}, {Giroletti}, {Yang}, {Marcote}, {Blanchard}, {Agudo}, {An}, {Bernardini}, {Beswick}, {Branchesi}, {Campana}, {Casadio}, {Chassande-Mottin}, {Colpi}, {Covino}, {D'Avanzo}, {D'Elia}, {Frey}, {Gawronski}, {Ghisellini}, {Gurvits}, {Jonker}, {van Langevelde}, {Melandri}, {Moldon}, {Nava}, {Perego}, {Perez-Torres}, {Reynolds}, {Salvaterra}, {Tagliaferri}, {Venturi}, {Vergani}, \& {Zhang}}]{ghirlanda2019}
{Ghirlanda}, G., {Salafia}, O.~S., {Paragi}, Z., {et~al.} 2019, Science, 363, 968

\bibitem[{{Giarratana} {et~al.}(2022){Giarratana}, {Rhodes}, {Marcote}, {Fender}, {Ghirlanda}, {Giroletti}, {Nava}, {Paredes}, {Ravasio}, {Rib{\'o}}, {Patel}, {Rastinejad}, {Schroeder}, {Fong}, {Gompertz}, {Levan}, \& {O'Brien}}]{Giarratana2022}
{Giarratana}, S., {Rhodes}, L., {Marcote}, B., {et~al.} 2022, \aap, 664, A36

\bibitem[{{Gill} \& {Granot}(2023)}]{gill2023}
{Gill}, R. \& {Granot}, J. 2023, \mnras, 524, L78

\bibitem[{{Gotz} {et~al.}(2022){Gotz}, {Mereghetti}, {Savchenko}, {Ferrigno}, {Bozzo}, \& {IBAS Team}}]{gotz2022}
{Gotz}, D., {Mereghetti}, S., {Savchenko}, V., {et~al.} 2022, GRB Coordinates Network, 32660, 1

\bibitem[{{Granot}(2008)}]{Granot2008}
{Granot}, J. 2008, \mnras, 390, L46

\bibitem[{{Granot} \& {Piran}(2012)}]{Granot2012}
{Granot}, J. \& {Piran}, T. 2012, \mnras, 421, 570

\bibitem[{{Granot} {et~al.}(1999){Granot}, {Piran}, \& {Sari}}]{Granot1999}
{Granot}, J., {Piran}, T., \& {Sari}, R. 1999, \apj, 513, 679

\bibitem[{{Granot} {et~al.}(2005){Granot}, {Ramirez-Ruiz}, \& {Loeb}}]{Granot2005_GRB030329}
{Granot}, J., {Ramirez-Ruiz}, E., \& {Loeb}, A. 2005, \apj, 618, 413

\bibitem[{{Greisen}(2003)}]{Greisen2003}
{Greisen}, E.~W. 2003, in Astrophysics and Space Science Library, Vol. 285, Information Handling in Astronomy - Historical Vistas, ed. A.~{Heck}, 109

\bibitem[{{Keimpema} {et~al.}(2015){Keimpema}, {Kettenis}, {Pogrebenko}, {Campbell}, {Cim{\'o}}, {Duev}, {Eldering}, {Kruithof}, {van Langevelde}, {Marchal}, {Molera Calv{\'e}s}, {Ozdemir}, {Paragi}, {Pidopryhora}, {Szomoru}, \& {Yang}}]{keimpema2015}
{Keimpema}, A., {Kettenis}, M.~M., {Pogrebenko}, S.~V., {et~al.} 2015, Experimental Astronomy, 39, 259

\bibitem[{{Kennea} {et~al.}(2022){Kennea}, {Williams}, \& {Swift Team}}]{kennea2022}
{Kennea}, J.~A., {Williams}, M., \& {Swift Team}. 2022, GRB Coordinates Network, 32635, 1

\bibitem[{{Kobayashi} {et~al.}(1999){Kobayashi}, {Piran}, \& {Sari}}]{Kobayashi1999}
{Kobayashi}, S., {Piran}, T., \& {Sari}, R. 1999, \apj, 513, 669

\bibitem[{{Kobayashi} \& {Sari}(2000)}]{Kobayashi2000}
{Kobayashi}, S. \& {Sari}, R. 2000, \apj, 542, 819

\bibitem[{{Kobayashi} \& {Zhang}(2003)}]{Kobayashi2003}
{Kobayashi}, S. \& {Zhang}, B. 2003, \apj, 597, 455

\bibitem[{{Kumar} \& {Granot}(2003)}]{Kumar2003}
{Kumar}, P. \& {Granot}, J. 2003, \apj, 591, 1075

\bibitem[{{Lapshov} {et~al.}(2022){Lapshov}, {Molkov}, {Mereminsky}, {Semena}, {Arefiev}, {Tkachenko}, {Lutovinov}, \& {SRG/ART-XC Team}}]{lapshov2022}
{Lapshov}, I., {Molkov}, S., {Mereminsky}, I., {et~al.} 2022, GRB Coordinates Network, 32663, 1

\bibitem[{{Laskar} {et~al.}(2023){Laskar}, {Alexander}, {Margutti}, {Eftekhari}, {Chornock}, {Berger}, {Cendes}, {Duerr}, {Perley}, {Ravasio}, {Yamazaki}, {Ayache}, {Barclay}, {Duran}, {Bhandari}, {Brethauer}, {Christy}, {Coppejans}, {Duffell}, {Fong}, {Gomboc}, {Guidorzi}, {Kennea}, {Kobayashi}, {Levan}, {Lobanov}, {Metzger}, {Ros}, {Schroeder}, \& {Williams}}]{laskar2023}
{Laskar}, T., {Alexander}, K.~D., {Margutti}, R., {et~al.} 2023, \apjl, 946, L23

\bibitem[{{Lesage} {et~al.}(2023){Lesage}, {Veres}, {Briggs}, {Goldstein}, {Kocevski}, {Burns}, {Wilson-Hodge}, {Bhat}, {Huppenkothen}, {Fryer}, {Hamburg}, {Racusin}, {Bissaldi}, {Cleveland}, {Dalessi}, {Fletcher}, {Giles}, {Hristov}, {Hui}, {Mailyan}, {Malacaria}, {Poolakkil}, {Roberts}, {von Kienlin}, {Wood}, {Ajello}, {Arimoto}, {Baldini}, {Ballet}, {Baring}, {Bastieri}, {Gonzalez}, {Bellazzini}, {Bissaldi}, {Blandford}, {Bonino}, {Bruel}, {Buson}, {Cameron}, {Caputo}, {Caraveo}, {Cavazzuti}, {Chiaro}, {Cibrario}, {Ciprini}, {Orestano}, {Crnogorcevic}, {Cuoco}, {Cutini}, {D'Ammando}, {De Gaetano}, {Di Lalla}, {Di Venere}, {Dom{\'\i}nguez}, {Fegan}, {Ferrara}, {Fleischhack}, {Fukazawa}, {Funk}, {Fusco}, {Galanti}, {Gammaldi}, {Gargano}, {Gasbarra}, {Gasparrini}, {Germani}, {Giacchino}, {Giglietto}, {Gill}, {Giroletti}, {Granot}, {Green}, {Grenier}, {Guiriec}, {Gustafsson}, {Hays}, {Hewitt}, {Horan}, {Hou}, {Kuss}, {Latronico}, {Laviron}, {Lemoine-Goumard}, {Li}, {Liodakis}, {Longo}, {Loparco}, {Lorusso},
  {Lovellette}, {Lubrano}, {Maldera}, {Manfreda}, {Mart{\'\i}-Devesa}, {Mazziotta}, {McEnery}, {Mereu}, {Meyer}, {Michelson}, {Mizuno}, {Monzani}, {Morselli}, {Moskalenko}, {Negro}, {Nuss}, {Omodei}, {Orlando}, {Ormes}, {Paneque}, {Panzarini}, {Persic}, {Pesce-Rollins}, {Pillera}, {Piron}, {Poon}, {Porter}, {Principe}, {Rain{\`o}}, {Rando}, {Rani}, {Razzano}, {Razzaque}, {Reimer}, {Reimer}, {Ryde}, {S{\'a}nchez-Conde}, {Parkinson}, {Scotton}, {Serini}, {Sgr{\`o}}, {Sharma}, {Siskind}, {Spandre}, {Spinelli}, {Tajima}, {Torres}, {Valverde}, {Venters}, {Wadiasingh}, {Wood}, \& {Zaharijas}}]{lesage2023}
{Lesage}, S., {Veres}, P., {Briggs}, M.~S., {et~al.} 2023, \apjl, 952, L42

\bibitem[{{Levan} {et~al.}(2023){Levan}, {Lamb}, {Schneider}, {Hjorth}, {Zafar}, {de Ugarte Postigo}, {Sargent}, {Mullally}, {Izzo}, {D'Avanzo}, {Burns}, {Ag{\"u}{\'\i} Fern{\'a}ndez}, {Barclay}, {Bernardini}, {Bhirombhakdi}, {Bremer}, {Brivio}, {Campana}, {Chrimes}, {D'Elia}, {Della Valle}, {De Pasquale}, {Ferro}, {Fong}, {Fruchter}, {Fynbo}, {Gaspari}, {Gompertz}, {Hartmann}, {Hedges}, {Heintz}, {Hotokezaka}, {Jakobsson}, {Kann}, {Kennea}, {Laskar}, {Le Floc'h}, {Malesani}, {Melandri}, {Metzger}, {Oates}, {Pian}, {Piranomonte}, {Pugliese}, {Racusin}, {Rastinejad}, {Ravasio}, {Rossi}, {Saccardi}, {Salvaterra}, {Sbarufatti}, {Starling}, {Tanvir}, {Th{\"o}ne}, {van der Horst}, {Vergani}, {Watson}, {Wiersema}, {Wijers}, \& {Xu}}]{levan2023}
{Levan}, A.~J., {Lamb}, G.~P., {Schneider}, B., {et~al.} 2023, \apjl, 946, L28

\bibitem[{{LHAASO Collaboration} {et~al.}(2023){LHAASO Collaboration}, {Cao}, {Aharonian}, {An}, {Axikegu}, {Bai}, {Bai}, {Bao}, {Bastieri}, {Bi}, \& et~al.}]{lhaaso2023}
{LHAASO Collaboration}, {Cao}, Z., {Aharonian}, F., {et~al.} 2023, Science, 380, 1390

\bibitem[{{Liu} {et~al.}(2022){Liu}, {Zhang}, {Xiong}, {Zheng}, {Wang}, {Xue}, {Qiao}, {Tan}, {Zhang}, {Li}, {Wen}, {Peng}, {Song}, {Zheng}, {Guo}, {Li}, {Ma}, {Huang}, {Zhao}, {Wang}, {Wang}, {Zhang}, {Du}, {Liang}, {Lu}, {Wu}, {Yu}, {Xiao}, {Cai}, {Zhang}, {Li}, {An}, {Gao}, {Gong}, {Liu}, {Liu}, {Sun}, {Xu}, {Yang}, {Feng}, {Wang}, {Zhang}, {Chen}, {Lu}, {Zhang}, {Gecam}, \& {Hebs Teams}}]{liu2022}
{Liu}, J.~C., {Zhang}, Y.~Q., {Xiong}, S.~L., {et~al.} 2022, GRB Coordinates Network, 32751, 1

\bibitem[{{Lyutikov}(2012)}]{Lyutikov2012}
{Lyutikov}, M. 2012, \mnras, 421, 522

\bibitem[{{Malesani} {et~al.}((in press)){Malesani}, {Levan}, {Izzo}, {de Ugarte Postigo}, {Ghirlanda}, {Heintz}, {Kann}, {Lamb}, {Palmerio}, {Salafia}, {Salvaterra}, {Tanvir}, {Ag{\"u}{\'\i} Fern{\'a}ndez}, {Campana}, {Chrimes}, {D'Avanzo}, {D'Elia}, {Della Valle}, {De Pasquale}, {Fynbo}, {Gaspari}, {Gompertz}, {Hartmann}, {Hjorth}, {Jakobsson}, {Palazzi}, {Pian}, {Pugliese}, {Ravasio}, {Rossi}, {Saccardi}, {Schady}, {Schneider}, {Sollerman}, {Starling}, {Th{\"o}ne}, {van der Horst}, {Vergani}, {Watson}, {Wiersema}, {Xu}, \& {Zafar}}]{malesani2023}
{Malesani}, D.~B., {Levan}, A.~J., {Izzo}, L., {et~al.} (in press), \aap, arXiv:2302.07891

\bibitem[{{Margutti} \& {Chornock}(2021)}]{margutti2021}
{Margutti}, R. \& {Chornock}, R. 2021, \araa, 59, 155

\bibitem[{{McMullin} {et~al.}(2007){McMullin}, {Waters}, {Schiebel}, {Young}, \& {Golap}}]{McMullin2007}
{McMullin}, J.~P., {Waters}, B., {Schiebel}, D., {Young}, W., \& {Golap}, K. 2007, in Astronomical Society of the Pacific Conference Series, Vol. 376, Astronomical Data Analysis Software and Systems XVI, ed. R.~A. {Shaw}, F.~{Hill}, \& D.~J. {Bell}, 127

\bibitem[{{Meszaros} \& {Rees}(1993)}]{meszaros1993}
{Meszaros}, P. \& {Rees}, M.~J. 1993, \apj, 405, 278

\bibitem[{{Mitchell} {et~al.}(2022){Mitchell}, {Phlips}, \& {Johnson}}]{mitchell2022}
{Mitchell}, L.~J., {Phlips}, B.~F., \& {Johnson}, W.~N. 2022, GRB Coordinates Network, 32746, 1

\bibitem[{{Mooley} {et~al.}(2018){Mooley}, {Deller}, {Gottlieb}, {Nakar}, {Hallinan}, {Bourke}, {Frail}, {Horesh}, {Corsi}, \& {Hotokezaka}}]{mooley2018}
{Mooley}, K.~P., {Deller}, A.~T., {Gottlieb}, O., {et~al.} 2018, \nat, 561, 355

\bibitem[{{Nappo} {et~al.}(2017){Nappo}, {Pescalli}, {Oganesyan}, {Ghirlanda}, {Giroletti}, {Melandri}, {Campana}, {Ghisellini}, {Salafia}, {D'Avanzo}, {Bernardini}, {Covino}, {Carretti}, {Celotti}, {D'Elia}, {Nava}, {Palazzi}, {Poppi}, {Prandoni}, {Righini}, {Rossi}, {Salvaterra}, {Tagliaferri}, {Testa}, {Venturi}, \& {Vergani}}]{nappo2017}
{Nappo}, F., {Pescalli}, A., {Oganesyan}, G., {et~al.} 2017, \aap, 598, A23

\bibitem[{{Natarajan} {et~al.}(2017){Natarajan}, {Paragi}, {Zwart}, {Perkins}, {Smirnov}, \& {van der Heyden}}]{Natarajan2017}
{Natarajan}, I., {Paragi}, Z., {Zwart}, J., {et~al.} 2017, \mnras, 464, 4306

\bibitem[{{O'Connor} {et~al.}(2023){O'Connor}, {Troja}, {Ryan}, {Beniamini}, {van Eerten}, {Granot}, {Dichiara}, {Ricci}, {Lipunov}, {Gillanders}, {Gill}, {Moss}, {Anand}, {Andreoni}, {Becerra}, {Buckley}, {Butler}, {Cenko}, {Chasovnikov}, {Durbak}, {Francile}, {Hammerstein}, {van der Horst}, {Kasliwal}, {Kouveliotou}, {Kutyrev}, {Lee}, {Srinivasaragavan}, {Topolev}, {Watson}, {Yang}, \& {Zhirkov}}]{oconnor2023}
{O'Connor}, B., {Troja}, E., {Ryan}, G., {et~al.} 2023, Science Advances, 9, eadi1405

\bibitem[{{Oren} {et~al.}(2004){Oren}, {Nakar}, \& {Piran}}]{Oren2004}
{Oren}, Y., {Nakar}, E., \& {Piran}, T. 2004, \mnras, 353, L35

\bibitem[{{Panaitescu} \& {Kumar}(2000)}]{Panaitescu2000}
{Panaitescu}, A. \& {Kumar}, P. 2000, \apj, 543, 66

\bibitem[{{Piano} {et~al.}(2022){Piano}, {Verrecchia}, {Bulgarelli}, {Ursi}, {Panebianco}, {Pittori}, {Longo}, {Parmiggiani}, {Tavani}, {Argan}, {Cardillo}, {Casentini}, {Evangelista}, {Foffano}, {Menegoni}, {Lucarelli}, {Addis}, {Baroncelli}, {di Piano}, {Fioretti}, {Fuschino}, {Romani}, {Marisaldi}, {Pilia}, {Trois}, {Donnarumma}, {Giuliani}, {Tempesta}, \& {Agile Team}}]{piano2022}
{Piano}, G., {Verrecchia}, F., {Bulgarelli}, A., {et~al.} 2022, GRB Coordinates Network, 32657, 1

\bibitem[{{Pihlstr{\"o}m} {et~al.}(2007){Pihlstr{\"o}m}, {Taylor}, {Granot}, \& {Doeleman}}]{Pihlstrom2007}
{Pihlstr{\"o}m}, Y.~M., {Taylor}, G.~B., {Granot}, J., \& {Doeleman}, S. 2007, \apj, 664, 411

\bibitem[{{Piran} {et~al.}(1993){Piran}, {Shemi}, \& {Narayan}}]{Piran1993}
{Piran}, T., {Shemi}, A., \& {Narayan}, R. 1993, \mnras, 263, 861

\bibitem[{{Planck Collaboration} {et~al.}(2020){Planck Collaboration}, {Aghanim}, {Akrami}, {Ashdown}, {Aumont}, {Baccigalupi}, {Ballardini}, {Banday}, {Barreiro}, {Bartolo}, \& et~al.}]{Planck2020}
{Planck Collaboration}, {Aghanim}, N., {Akrami}, Y., {et~al.} 2020, \aap, 641, A6

\bibitem[{{Ren} {et~al.}(2024){Ren}, {Wang}, \& {Dai}}]{Ren2023b}
{Ren}, J., {Wang}, Y., \& {Dai}, Z.-G. 2024, \apj, 962, 115

\bibitem[{{Ren} {et~al.}(2023){Ren}, {Wang}, {Zhang}, \& {Dai}}]{ren2023}
{Ren}, J., {Wang}, Y., {Zhang}, L.-L., \& {Dai}, Z.-G. 2023, \apj, 947, 53

\bibitem[{{Rhoads}(1997)}]{Rhoads1997}
{Rhoads}, J.~E. 1997, \apjl, 487, L1

\bibitem[{{Ripa} {et~al.}(2022){Ripa}, {Pal}, {Werner}, {Ohno}, {Takahashi}, {Meszaros}, {Csak}, {Dafcikova}, {Munz}, {Husarikova}, {Breuer}, {Topinka}, {Hroch}, {Urbanec}, {Kasal}, {Povalac}, {Hudec}, {Kapus}, {Frajt}, {Laszlo}, {Koleda}, {Smelko}, {Hanak}, {Lipovsky}, {Galgoczi}, {Uchida}, {Poon}, {Matake}, {Uchida}, {Bozoki}, {Dalya}, {Enoto}, {Frei}, {Friss}, {Fukazawa}, {Hirose}, {Hisadomi}, {Ichinohe}, {Kapas}, {Kiss}, {Mizuno}, {Nakazawa}, {Odaka}, {Takatsy}, {Torigoe}, {Kogiso}, {Yoneyama}, {Moritaki}, {Kano}, \& {GRBAlpha Collaboration.}}]{ripa2022}
{Ripa}, J., {Pal}, A., {Werner}, N., {et~al.} 2022, GRB Coordinates Network, 32685, 1

\bibitem[{{Salafia} {et~al.}(2022){Salafia}, {Ravasio}, {Yang}, {An}, {Orienti}, {Ghirlanda}, {Nava}, {Giroletti}, {Mohan}, {Spinelli}, {Zhang}, {Marcote}, {Cim{\`o}}, {Wu}, \& {Li}}]{Salafia2022_29A}
{Salafia}, O.~S., {Ravasio}, M.~E., {Yang}, J., {et~al.} 2022, \apjl, 931, L19

\bibitem[{{Sari} \& {Piran}(1995)}]{Sari1995}
{Sari}, R. \& {Piran}, T. 1995, \apjl, 455, L143

\bibitem[{{Sato} {et~al.}(2023){Sato}, {Murase}, {Ohira}, \& {Yamazaki}}]{sato2023}
{Sato}, Y., {Murase}, K., {Ohira}, Y., \& {Yamazaki}, R. 2023, \mnras, 522, L56

\bibitem[{{Shepherd} {et~al.}(1994){Shepherd}, {Pearson}, \& {Taylor}}]{Shepherd1994}
{Shepherd}, M.~C., {Pearson}, T.~J., \& {Taylor}, G.~B. 1994, in Bulletin of the American Astronomical Society, Vol.~26, 987--989

\bibitem[{{Tan} {et~al.}(2022){Tan}, {Li}, {Ge}, {Li}, {Xiong}, \& {Zhang}}]{tan2022}
{Tan}, W.~J., {Li}, C.~K., {Ge}, M.~Y., {et~al.} 2022, The Astronomer's Telegram, 15660, 1

\bibitem[{{Taylor} {et~al.}(2004){Taylor}, {Frail}, {Berger}, \& {Kulkarni}}]{taylor2004}
{Taylor}, G.~B., {Frail}, D.~A., {Berger}, E., \& {Kulkarni}, S.~R. 2004, \apjl, 609, L1

\bibitem[{{Taylor} {et~al.}(2005){Taylor}, {Momjian}, {Pihlstr{\"o}m}, {Ghosh}, \& {Salter}}]{Taylor2005}
{Taylor}, G.~B., {Momjian}, E., {Pihlstr{\"o}m}, Y., {Ghosh}, T., \& {Salter}, C. 2005, \apj, 622, 986

\bibitem[{{Thompson} {et~al.}(2017){Thompson}, {Moran}, \& {Swenson}}]{Thompson2017}
{Thompson}, A.~R., {Moran}, J.~M., \& {Swenson}, George~W., J. 2017, {Interferometry and Synthesis in Radio Astronomy, 3rd Edition} (Springer Cham)

\bibitem[{{Ursi} {et~al.}(2022){Ursi}, {Panebianco}, {Pittori}, {Verrecchia}, {Longo}, {Parmiggiani}, {Tavani}, {Argan}, {Cardillo}, {Casentini}, {Evangelista}, {Foffano}, {Menegoni}, {Piano}, {Lucarelli}, {Addis}, {Baroncelli}, {Bulgarelli}, {di Piano}, {Fioretti}, {Fuschino}, {Romani}, {Marisaldi}, {Pilia}, {Trois}, {Donnarumma}, {Giuliani}, {Tempesta}, \& {Agile Team}}]{ursi2022}
{Ursi}, A., {Panebianco}, G., {Pittori}, C., {et~al.} 2022, GRB Coordinates Network, 32650, 1

\bibitem[{{van Eerten} {et~al.}(2010){van Eerten}, {Zhang}, \& {MacFadyen}}]{vanEerten2010}
{van Eerten}, H., {Zhang}, W., \& {MacFadyen}, A. 2010, \apj, 722, 235

\bibitem[{{Veres} {et~al.}(2022){Veres}, {Burns}, {Bissaldi}, {Lesage}, {Roberts}, \& {Fermi GBM Team}}]{veres2022}
{Veres}, P., {Burns}, E., {Bissaldi}, E., {et~al.} 2022, GRB Coordinates Network, 32636

\bibitem[{{Xiao} {et~al.}(2022){Xiao}, {Krucker}, \& {Daniel}}]{xiao2022}
{Xiao}, H., {Krucker}, S., \& {Daniel}, R. 2022, GRB Coordinates Network, 32661, 1

\bibitem[{{Yi} {et~al.}(2013){Yi}, {Wu}, \& {Dai}}]{Yi2013}
{Yi}, S.-X., {Wu}, X.-F., \& {Dai}, Z.-G. 2013, \apj, 776, 120

\bibitem[{{Zheng} {et~al.}(2024){Zheng}, {Wang}, {Liu}, \& {Zhang}}]{Zheng2024}
{Zheng}, J.-H., {Wang}, X.-Y., {Liu}, R.-Y., \& {Zhang}, B. 2024, \apj, 966, 141

\bibitem[{{Zou} {et~al.}(2005){Zou}, {Wu}, \& {Dai}}]{zou2005}
{Zou}, Y.~C., {Wu}, X.~F., \& {Dai}, Z.~G. 2005, \mnras, 363, 93

\end{thebibliography}

\begin{appendix}
\section{EVN observation strategy and data reduction}
\label{appendix1}

\begin{table*}[t]
\caption[]{List of antennas that join each observing run.}
\label{tab:antennas}
\centering

\begin{tabular}{lrcl}
\toprule
Code    &$t_\mathrm{obs}-t_0$  &Array  &Antennas\\ 
 &[days]    &  &\\ 
\midrule
RG013 B &40   &EVN  &Wb, Ef, Nt, O6, Ur, Tm, Ys, Tr, Hh, Mh\\
RG013 C &43   &EVN  &Jb, Wb, Ef, Mc, O8, Ur, Tm, Ys, Tr, Hh\\
BA160 B &44  &VLBA  &Fd, Hn, Mk, Nl, Ov, Pt, Sc\\ 
BA160 C &114 &VLBA  &Br, Fd, La, Mk, Nl, Ov, Pt, Sc\\
RG013 D &117  &EVN  &Jb, Wb, Ef, Mc, O8, Ur, Tm, Ys, Tr\\
RG013 E &118  &EVN  &Wb, Ef, Mc, Nt, O6, Ur, Tm, Ys, Tr, Hh, Mh\\
BA160 C1 &205 &VLBA  &Br, Fd, Hn, Kp, La, Mk, Nl, Ov, Pt, Sc\\
RG013 F &261  &EVN  &Jb, Ef, Mc, Nt, O8, Tm, Ys, Tr, Hh, Ir\\
BA160 D &262 &VLBA  &Br, Fd, Hn, Kp, La, Mk, Nl, Ov, Pt, Sc\\

\bottomrule
\end{tabular}

\vspace{5pt}
Wb: Westerbork, 25m; Ef: Effelsberg, 100m; Nt: Noto, 32m; O6: Onsala, 20m; Ur: Urumqi; Tm: Tianma, 65m; Ys: Yebes, 40m; Tr: Torun, 32m; Hh: Hartebeesthoek, 25m; Mh: Mets\"ahovi, 14m; Jb: Jodrell bank (Lovell), 76m; Mc: Medicina, 32m; O8: Onsala, 25m; Ir: Irbene; Br: Brewster, 25m; Fd: Fort Davis, 25m; Hn: Hancock, 25m; Kp: Kitt Peak, 25m; La: Los Alamos, 25m; Mk: Mauna Kea, 25m; Nl: North Liberty, 25m; Ov: Owen Valley, 25m; Pt: Pie Town, 25m; Sc: Saint Croix, 25m.
\end{table*}
In this Appendix we provide detailed information on the observation strategy and the data reduction of the EVN observations. As explained in Section \ref{sec:observations}, the structure of the observations followed a typical phase-referencing experiment. Three compact, extragalactic radio sources J1800$+$3848, J1925$+$2106 and J0121$+$0422 were used as fringe finders and bandpass calibrators in the campaign. The target scans, lasting approximately 4.5 and 2.5\, minutes at 5 and 8 GHz, respectively, were interleaved with 1.5\,minute scans of the phase calibrator. In the first two observations, namely 40 and 43 days post-burst (RG013 B and C), the radio source J1905$+$1943 was used as a phase calibrator and the Very Large Array Sky Survey (VLASS) compact radio source J191142$+$1952 was included for testing its suitability as a closer phase reference source ($d=0.33^{\circ}$ from the GRB position).  Given the success of that test, J191142$+$1952 was then adopted as phase calibrator in the last three epochs, i.e. from 117 to 261 days post-burst (RG013 D, E and F). In order to inspect the consistency of the calibration procedure, one or multiple compact radio sources were observed approximately every 30\,minutes. The phase and amplitude solutions derived from the calibrators were applied to these check sources to verify the quality of the calibration.

The calibration was performed using {\tt AIPS} \citep{Greisen2003}, following the standard procedure for EVN phase-referenced observations. The amplitude calibration, which accounts for the bandpass response, the antenna gain curves and the system temperatures, was performed using the results from the EVN pipeline. Procedures \texttt{vlbatecr} and \texttt{vlbampcl} were used to correct for the dispersive delay and to calculate the manual single band delay on the fringe finder, respectively. Subsequently, we carried out the global fringe fitting on the phase calibrator with the task \texttt{fring}. Solutions were interpolated and applied to the phase calibrator itself, the check sources and the target with the task \texttt{clcal}. After this point, the calibration procedure differs according to the epoch.

For the first epoch (RG013 B), we carried out the fringe fitting on J1905$+$1943 using a model of the source derived by a concatenation (in {\tt CASA}, \citealt{McMullin2007}) and self-calibration (in {\tt Difmap}) of all the visibilities on the source obtained across the two epochs at 8\,GHz. This approach is warranted by the stability of the structure of extragalactic sources on the duration of the campaign, and improves the phase, delay, and rate calibration by accounting for the possibile structure of the phase-reference source. 
We then interpolated the solutions and applied the results to J1905$+$1943 itself, the check source (J1923$+$2010) and GRB\,221009A. Moreover, for RG013 B and C, we tried two rounds of self-calibration on J1905$+$1943, first in phase-only, with a solution interval of 2\,minutes, and then in amplitude and phase with a solution interval of 60\, minutes in {\tt AIPS}. 
Since the self-calibration on J1905$+$1943 did not result in a significant improvement in the final S/R of the image of the GRB, we used the image generated with the calibration performed prior to the self-calibration.

In the last three epochs, we employed a different phase calibrator, J191142$+$1952, motivated by the significantly smaller separation from the source (0.33$^{\circ}$ vs 1.75$^{\circ}$). If the phase calibrator is closer to the target, any possible decorrelation of the phase solutions is significantly reduced. However, the position of J191142$+$1952 was constrained only with a precision of the order of an arcsecond: for VLBI observations, this means that the coordinates of the centre of the source were not aligned with the phase centre of the observation. If one does not correct for the uncertainty in the position, the phase solutions of the global fringe fitting on the phase calibrator will contain a systematic error and the apparent coordinates of the centre of the sources to which these solutions are applied will be incorrect. To avoid this, we started from the fourth epoch, made at 8 GHz, which provides higher angular resolution and therefore a more precise position of the calibrator. We applied the solutions of the first global fringe fitting on J191142$+$1952 to the check sources, J1905$+$1943 and J1923$+$2010, we produced an image of each of them and we compared the apparent coordinates of each source with the actual position, known with an uncertainty of the order of mas. Since J1905$+$1943 and J1923$+$2010 appear to be aligned in the sky, with J191142$+$1952 placed in between, we derived the real coordinates with a 1D interpolation at the position of J191142$+$1952 of the offset observed for the check sources. We then re-calculated the visibilities of J191142$+$1952 by fixing the phase centre with the \texttt{fixvis} task in {\tt CASA}, inserting the new coordinates. We then repeated the entire calibration process iteratively, until the apparent and the real sky coordinates of the check sources were consistent within the resolution of the observation. We corrected the third and fifth epoch using the position derived at 8.3\,GHz. 

Subsequently, we produced a model of J191142$+$1952 in {\tt Difmap} and we used the model as input to perform the global fringe fitting of the phase calibrator in {\tt AIPS}, in order to take into account any possible structure of J191142$+$1952 and correct for it. We interpolated the solutions and we applied them to J191142$+$1952, J1905$+$1943, J1923$+$2010 and GRB\,221009A. Lastly, we performed a round of amplitude and phase self-calibration of J191142$+$1952 in {\tt AIPS}, using a solution interval of 2\,minutes and we applied the interpolated solutions to J191142$+$1952, J1905$+$1943, J1923$+$2010 and GRB\,221009A. After each of the aforementioned steps of the procedure, the derived solutions were inspected and bad data were properly flagged. 

Images of the target and the check sources were produced using {\tt Difmap}. 
For the RG013 C dataset, both the noise pattern in the dirty image and the modelfit residuals indicated the presence of some residual phase calibration errors. Of all our datasets, this is the epoch with the largest flux density ($S=1.76$ mJy, as estimated interpolating the VLA measurments reported by \citealt{laskar2023}).  Therefore, considering the presence of the sensitive Effelsberg and Tianma65 radio telescopes and the large bandwidth, we carried out a phase-only self calibration with a solution time interval of one minute and combining all the frequency sub-bands. After carefully checking the behaviour of the solutions, we concluded that the process  improved the data quality without introducing any artifact in the data. We carried out the subsequent steps of analysis as for the other epochs.

\section{Circular Gaussian fits to source visibilities}\label{apx:corner_plots_circ_gauss_fits}

In this appendix we provide more detailed information about our circular Gaussian fits to source visibilities from our VLBI observations. Our observations and calibration procedure yielded a set of $N$ complex visibilities $\lbrace\mathcal{V}_i\rbrace_{i=1}^N$, which measure the two-dimensional Fourier transform of the image at positions $(u_i,v_i)$ in the $(u,v)$ plane, with associated uncertainties $\sigma_i=w_i^{-1/2}$, where $w_i$ are the `data weights' determined by the calibration procedure. Each visibility resulted from interferometry between the pair of antennae (`baseline') $B_i=(b_{1,i},b_{2,i})$, where $b_{1,i}$ and $b_{2,i}$ are indices that identify the two antennae, which satisfy $1\leq b_{1,i} < N_\mathrm{A}$ and $b_{1,i}<b_{2,i}\leq N_\mathrm{A}$, were $N_\mathrm{A}$ is the total number of antennae in the VLBI network. The total number of distinct baselines is $N_\mathrm{B}=N_\mathrm{A}(N_\mathrm{A}-1)/2$. 

We assumed a Gaussian likelihood $\mathcal{L}$ for these visibilities, 
\begin{equation}
    \ln\mathcal{L}\left(\lbrace\mathcal{V}_i\rbrace_{i=1}^{N}\,|\,\vec x\right) = -\frac{1}{2} \sum_{i=0}^{N}\frac{\left(\mathcal{V}_{\mathrm{m}}(u_i,v_i,\vec x)-\mathcal{V}_{i}\right)^2}{\sigma_{i}^2},
\end{equation}
where $\vec x$ is the model parameter vector and $\mathcal{V}_\mathrm{m}(u_i,v_i,\vec x)$ is the visibility predicted by the model. We adopted a circular Gaussian surface brightness model, with amplitude correction factors introduced to account for potential antenna-based systematic mis-calibrations \citep[similar to][]{Natarajan2017}, namely 
\begin{equation}
    \mathcal{V}_\mathrm{m}(u_i,v_i,\vec x) = G_{b_{1,i}}^{-1}G_{b_{2,i}}^{-1} F_\nu e^{-2\pi^2 \left(\frac{s}{\sqrt{8\ln 2}}\right)^2\left(u_i^2+v_i^2\right)-2\pi j (u_i\rho + v_i\delta)},
\end{equation}
where $j=\sqrt{-1}$, $F_\nu$ is the total flux density, $s$ is the full width at half maximum of the circular Gaussian, $\rho$ and $\delta$ are the spherical offsets of the source with respect to the phase centre, and $G_{b_{1,i}}$ and $G_{b_{2,i}}$ are dimensionless factors that encode the amplitude mis-calibrations of antennae $b_{1,i}$ and $b_{2,i}$ \citep{Natarajan2017}. These are defined as follows: $G_{i}=1$ means that there is no systematic error in the amplitude calibration of antenna $i$; conversely, $G_{i}=0.9$ (resp.\ $G_{i}=1.1)$ means that the amplitude of that particular antenna is systematically overestimated (resp.\ underestimated) by 10\%.

The parameter vector of the model is therefore $\vec x = (F_\nu,s,\rho,\delta,G_1,...,G_{N_\mathrm{A}})$. By Bayes' theorem, we defined the posterior  probability on $\vec x$, given our data $\lbrace\mathcal{V}_i\rbrace_{i=1}^{N}$, as
\begin{equation}
    P\left(\vec x\,|\,\lbrace\mathcal{V}_i\rbrace_{i=1}^{N}\right) \propto \pi(\vec x)\mathcal{L}\left(\lbrace\mathcal{V}_i\rbrace_{i=1}^{N}\,|\,\vec x\right), 
    \label{eq:circ_gauss_posterior}
\end{equation}
where $\pi(\vec x)$ is the prior probability on the parameters. We decomposed the latter into a product of independent one-dimensional priors on all parameters. We adopted simple independent uniform priors on the source parameters, with the due bounds $F_\nu>0$ and $s>0$. Where necessary, in order to prevent the fitting procedure from picking up some noise peak instead of the actual source, we restricted the position $(\rho,\delta)$ to within a small angular distance $\Delta_\mathrm{pos}\sim 1\,\mathrm{mas}$ from the peak $(\rho_0,\delta_0)$ of the cleaned map constructed with {\tt Difmap}. All, but one, of the priors on the mis-calibration parameters $G_{i}$ were set equal to Gaussians centered at 1 with sigma equal to 0.1, hence admitting typical systematic errors of 10\%. The prior on the remaining $G_{i}$ parameter was set to a delta function centered at 1, in order to break a degeneracy with the total flux density (see, e.g., \citealt{Natarajan2017}). The index $i$ of such fixed calibration parameter corresponds to the reference antenna in the network, which we chose to be Effelsberg for the EVN and Fort Davis for the VLBA (with the exception of epoch BA160C1, where we used Los Alamos).

For each epoch, we sampled the posterior probability using the
\textsc{emcee} \citep{Foreman-Mackey2013} python package. We initialised \textsc{emcee} with the initial guess $\vec x = (I_{\nu,\mathrm{pk}},10^{-2}\,\mathrm{mas},\rho_0,\delta_0,1,...,1)$, where $I_{\nu,\mathrm{pk}}$ is the peak surface brightness (expressed in Jy/beam) in the cleaned map, corresponding to an unresolved circular Gaussian source at the position of the peak of the cleaned map and with a flux density that yields the observed peak surface brightness.  We then ran $5000$ iterations of the MCMC (depending on the epoch) with 32 walkers, producing $\sim 10^5$ samples of the posterior probability density for each epoch. The results were constructed after discarding the initial 30\% of these samples in each chain as burn-in. 

\begin{figure*}
\centering
\includegraphics[width=0.33\textwidth]{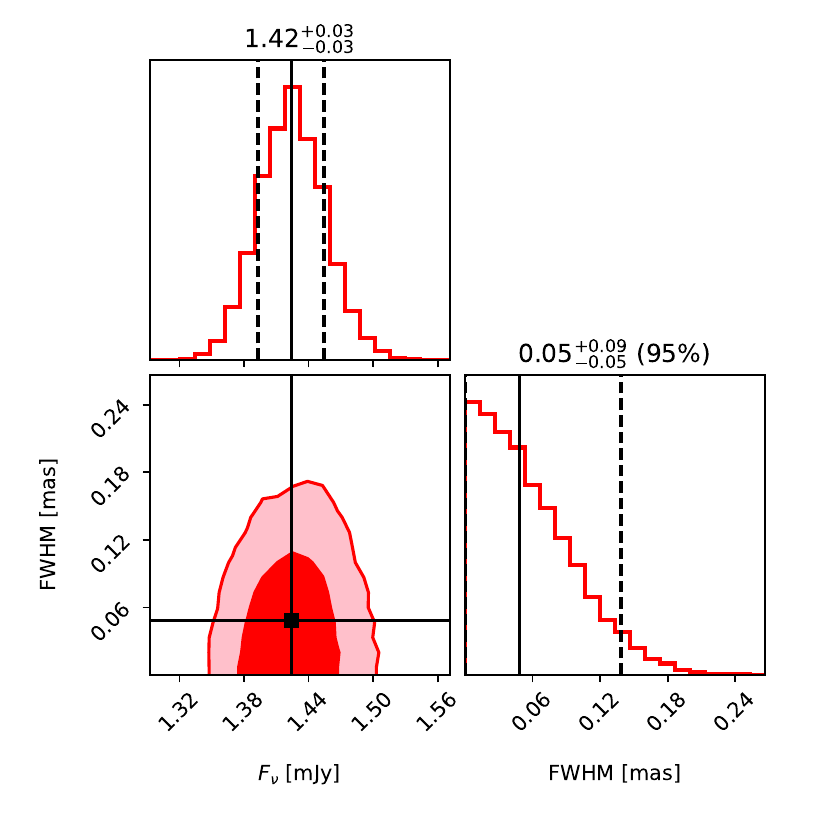}%
\includegraphics[width=0.33\textwidth]{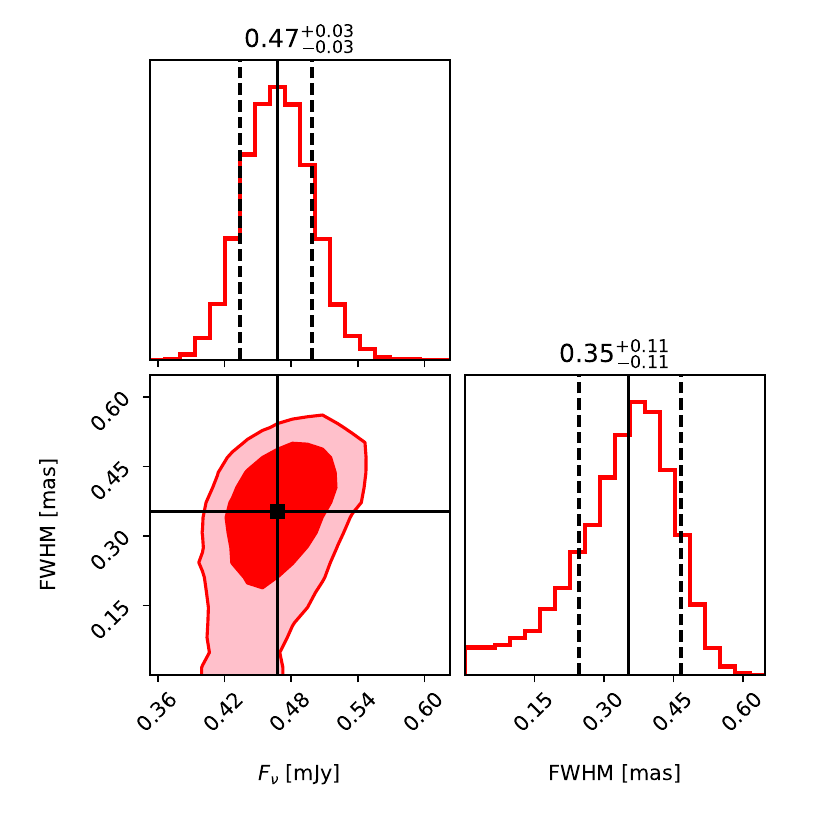}%
\includegraphics[width=0.33\textwidth]{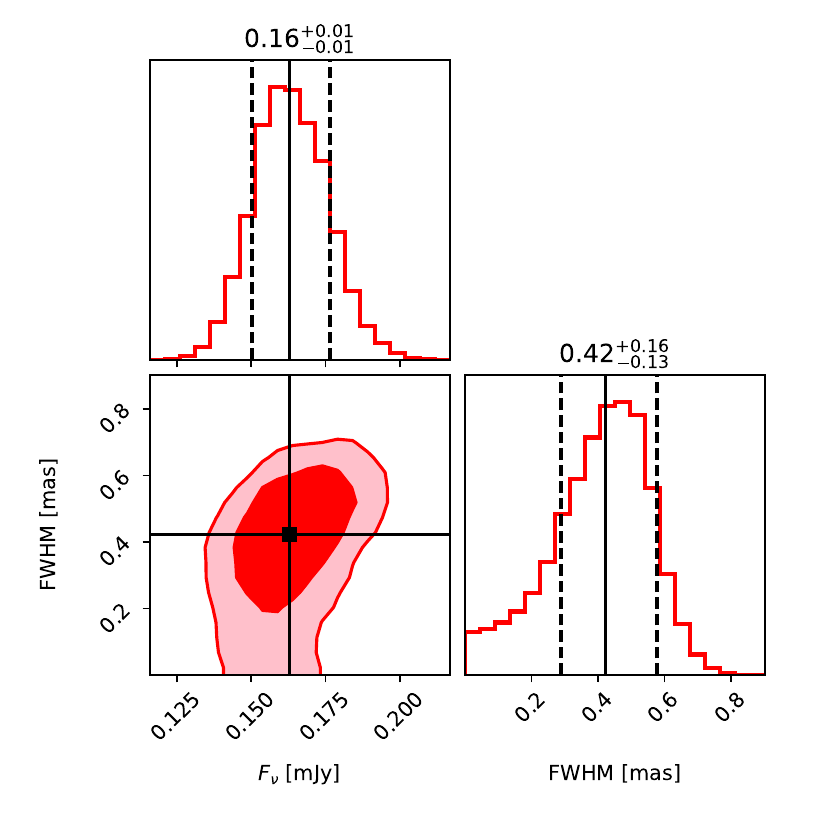}
\caption{Posterior probability distribution on the source size and flux density in our EVN 4.9\,GHz epochs at $T-T_0=43$ (RG1013 C, left), 117 (RG013 D, centre) and 261 (RG013 F, right) days. In each corner plot, the top-left and bottom-right sub-panels show histograms of the posterior samples of $F_\nu$ and FWHM, with the vertical solid lines showing the median and the vertical dashed lines bracketing the 68\% credible interval or, if the latter extends to 0, the 95\% credible upper limit. The bottom-left sub-panel of each corner plot shows the smallest contours containing 68\% and 95\% of the posterior probability.}
    \label{fig:EVN_5GHz_corner}
\end{figure*}

\begin{figure*}
\sidecaption
\begin{minipage}{0.66\textwidth}
\includegraphics[width=0.5\textwidth]{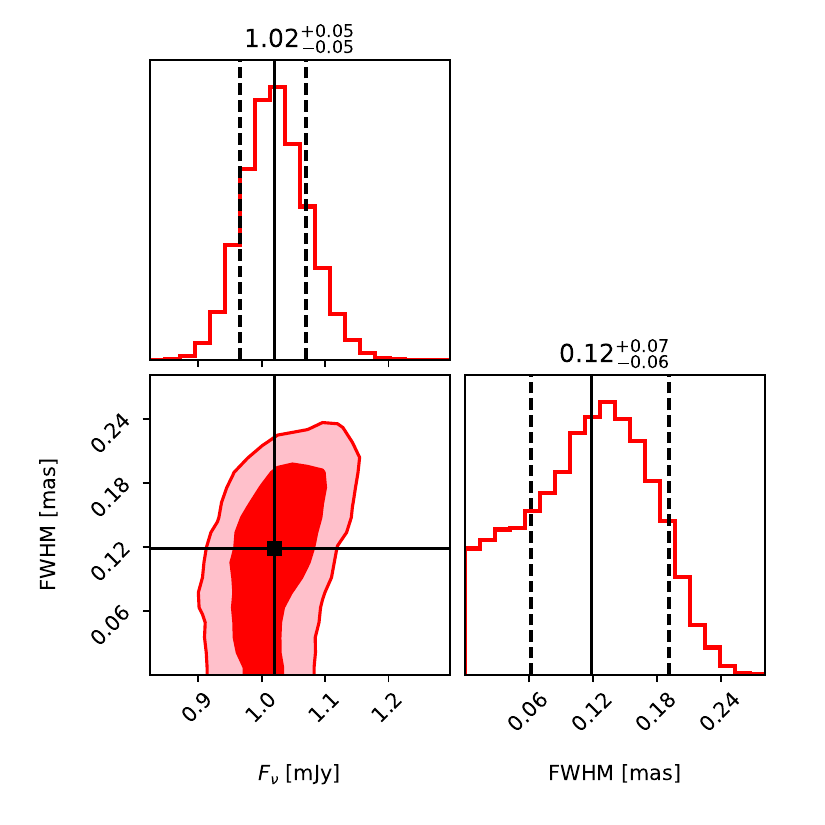}
\includegraphics[width=0.5\textwidth]{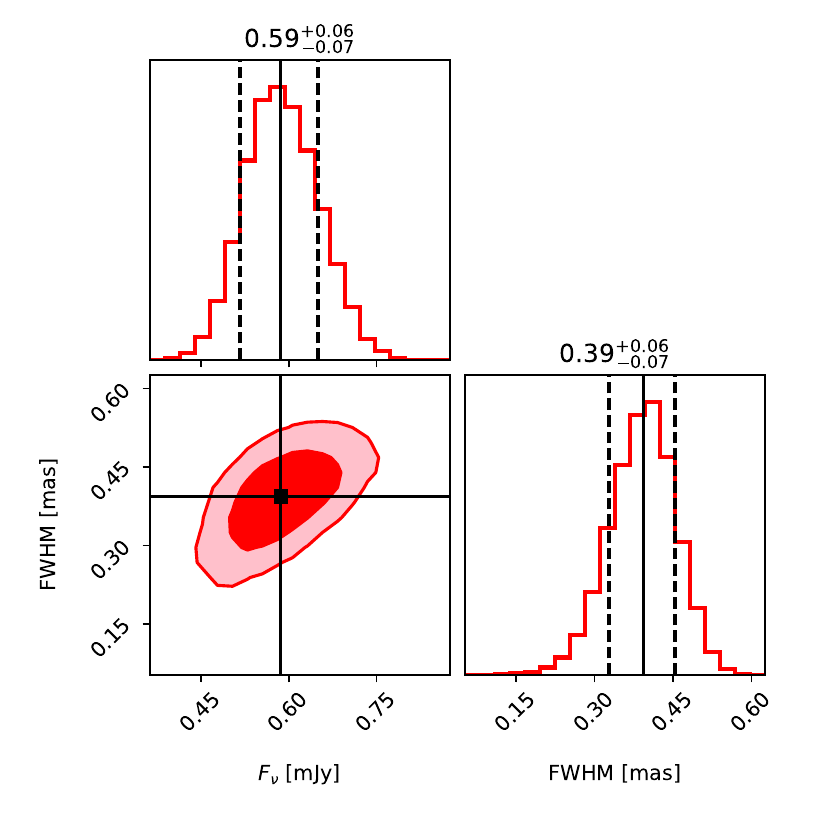}    
\end{minipage}
\caption{Similar to Figure \ref{fig:EVN_5GHz_corner}, but for our EVN 8.3\,GHz epochs at $T-T_0=40$ days (EVN B, left) and 118 days (EVN E, right).}
\label{fig:EVN_8GHz_corner}
\end{figure*}

\begin{figure*}
\sidecaption
\begin{minipage}{0.66\textwidth}
\includegraphics[width=0.5\textwidth]{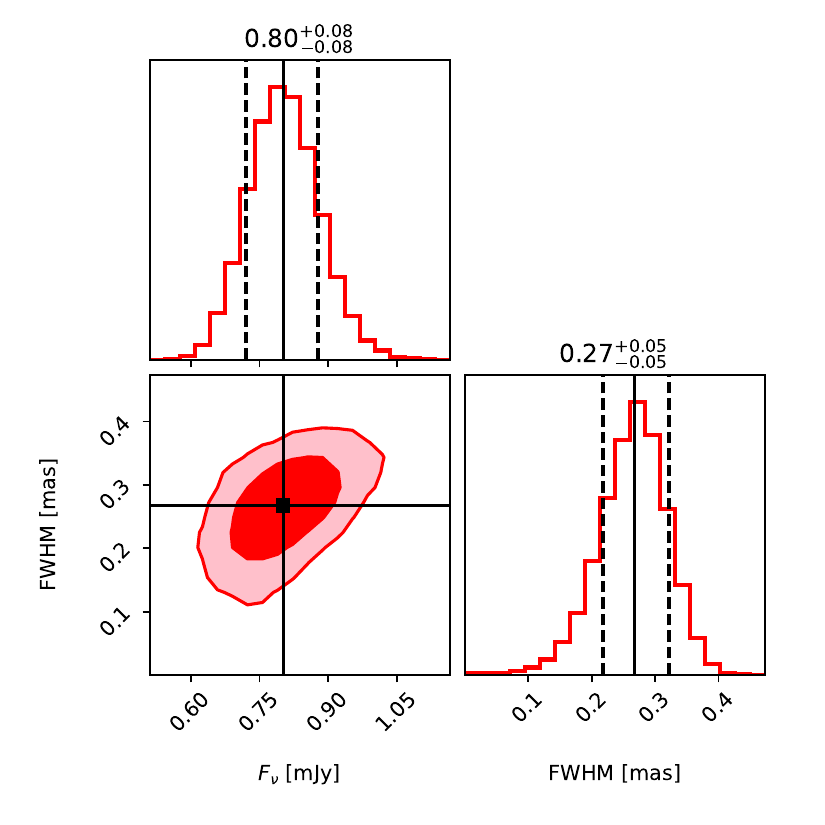}%
\includegraphics[width=0.5\textwidth]{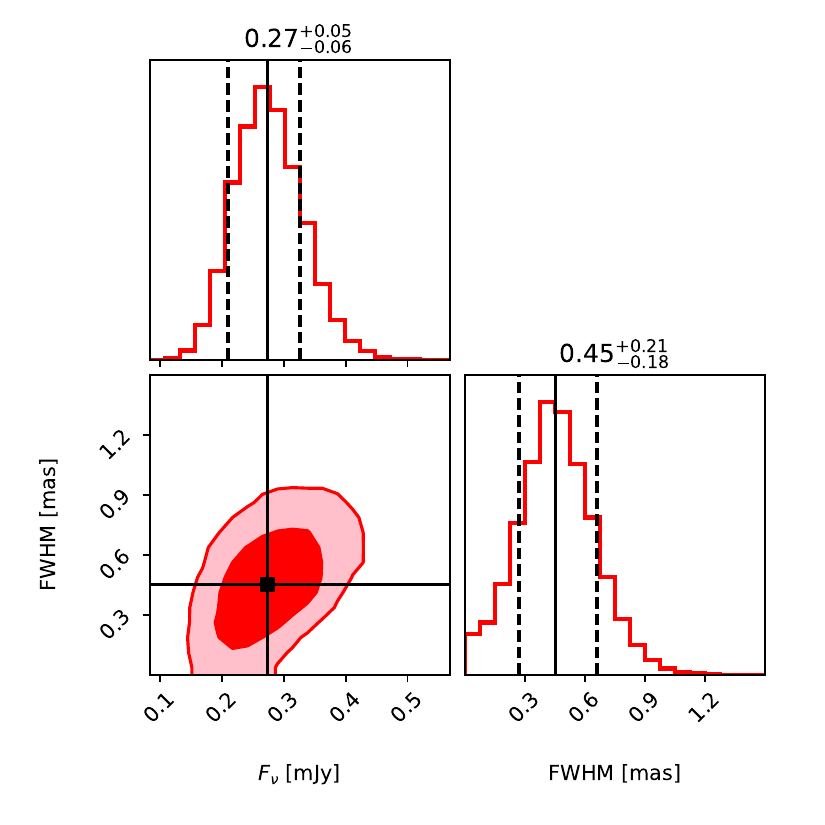}\\
\includegraphics[width=0.5\textwidth]{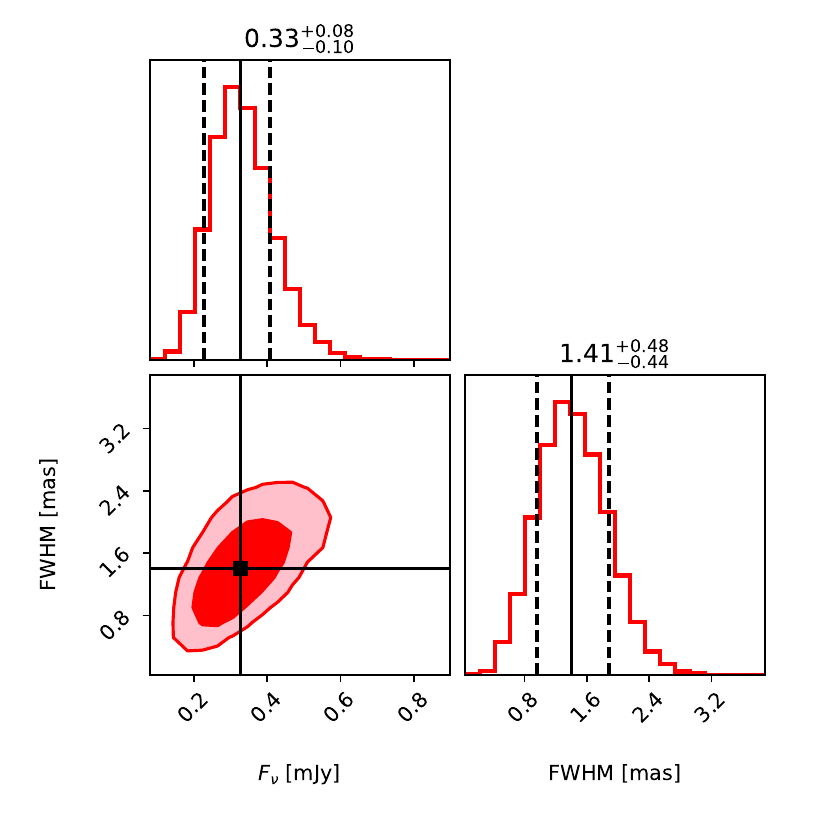}%
\includegraphics[width=0.5\textwidth]{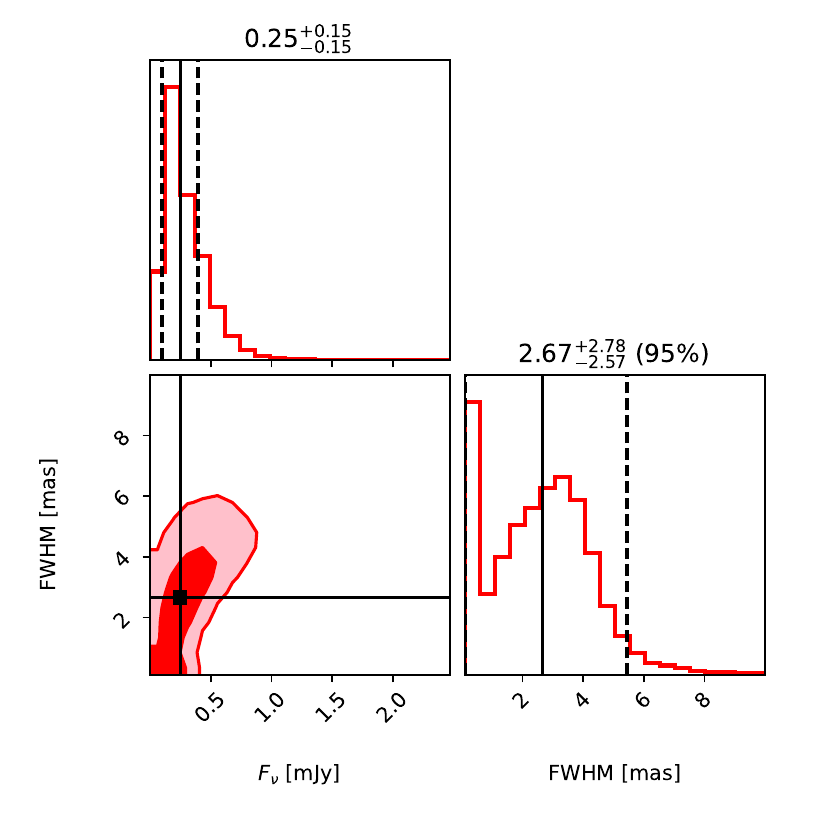}
\end{minipage}
\caption{Similar to Figure \ref{fig:EVN_5GHz_corner}, but for our VLBA 15\,GHz epochs at $T-T_0=44$ (VLBA B, upper left), 114 (VLBA C, upper right), 205 (VLBA C1, lower left) and 262 (VLBA D, lower right) days.}
\label{fig:VLBA_15GHz_corner}
\end{figure*}

Figure \ref{fig:EVN_5GHz_corner} shows corner plots that illustrate the properties of the posterior probability density of $(F_\nu,s)$ for our EVN 4.9\,GHz epochs. Figures \ref{fig:EVN_8GHz_corner} and \ref{fig:VLBA_15GHz_corner} show the corresponding corner plots for our EVN 8.3\,GHz epochs and for our VLBA 15\,GHz epochs, respectively.

\section{Tests on the evolution of the flux density and the size}
\label{appendix:tests}

\begin{figure}
    \centering
    \includegraphics[width=\columnwidth]{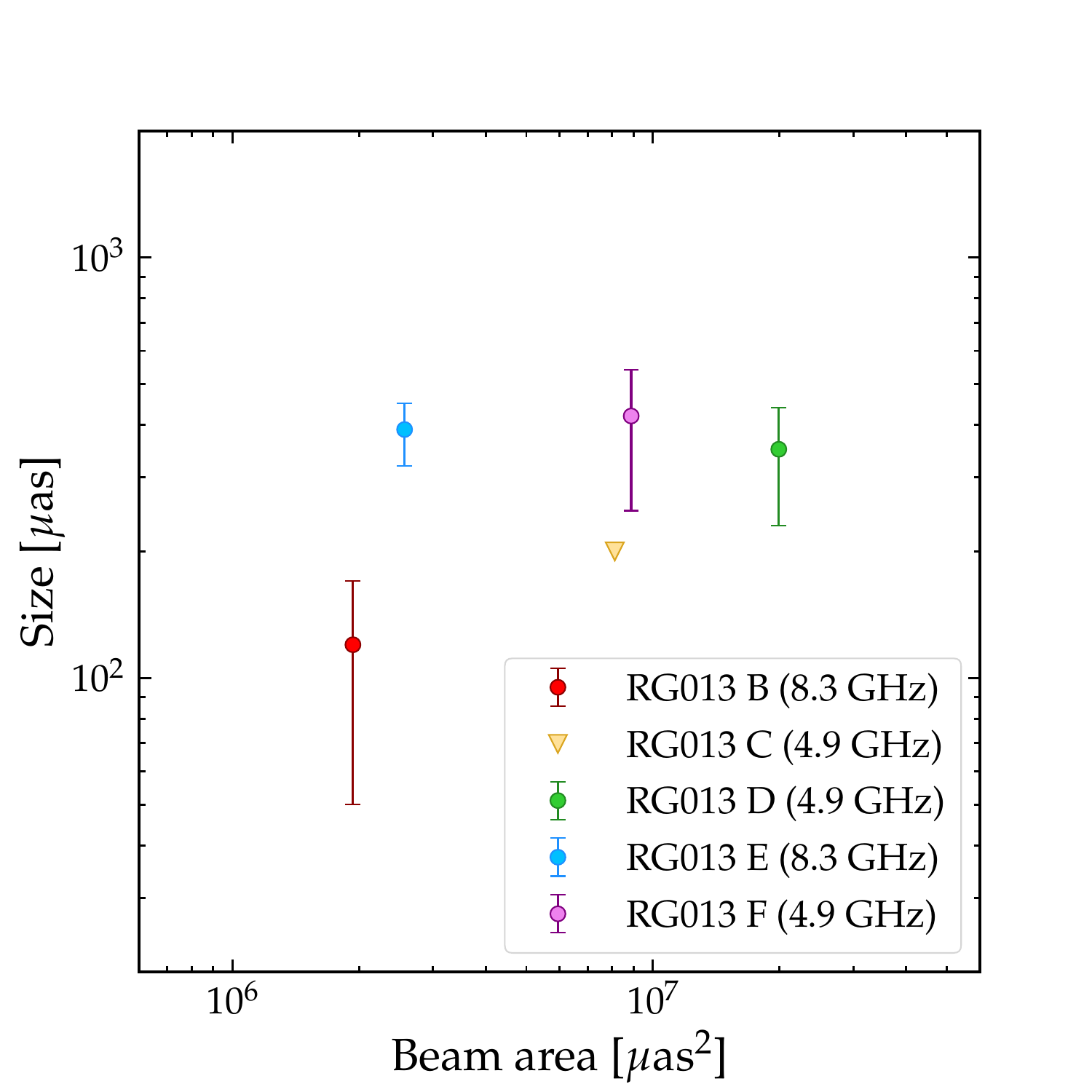}
    \caption{Estimates of the FWHM of the GRB as a function of the total area of the synthesised beam for the EVN observations.}
    \label{fig:GRB_size_vs_Beam}
\end{figure}

\begin{figure*}
\sidecaption
\centering
\begin{minipage}{0.66\textwidth}
\includegraphics[width=\textwidth]{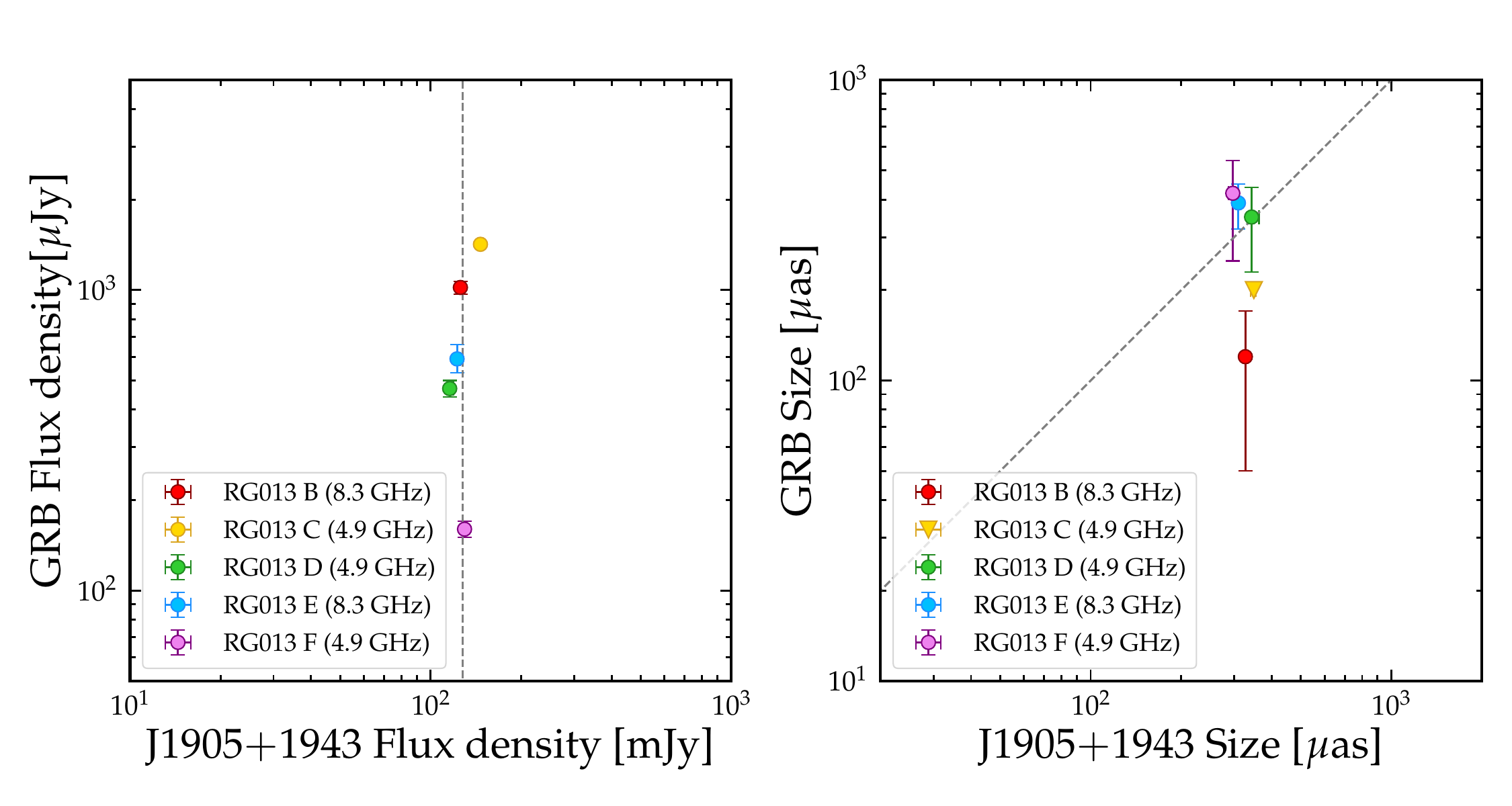}
\end{minipage}
\caption{Flux density (left panel) and size (right panel) of GRB\,221009A and J1905$+$1943 in the EVN observations. The average flux density of J1905$+$1943 (left panel) and the 1:1 correlation (right panel) are shown as grey dashed lines for the sake of comparison.}
    \label{fig:GRB_vs_J1905_EVN}
\end{figure*}

In this Appendix we present tests on the EVN observational results that we carried out in order to exclude the possibility that the measured evolution of the GRB size is a result of systematic effects. These tests include the check source J1905$+$1943. Unfortunately, due to the sparse $(u,v)$ plane coverage and the large separation, J1923$+$2010 could not be used to get meaningful constraints.
First, the measured GRB afterglow size as a function of the area of the synthesised beam are presented in Fig.\ \ref{fig:GRB_size_vs_Beam}. These quantities are clearly not correlated, hence we can exclude the possibility that the observed expansion of the GRB is driven by a systematic change in the width of the synthesised beam. In Fig.\ \ref{fig:GRB_vs_J1905_EVN}, the flux density (left panel) and the size (right panel) of GRB\,221009A and the check source J1905$+$1943 are compared. The decrease in the GRB\,221009A flux density is not accompanied by a variation of the J1905$+$1943 flux density, as expected. Regarding the size measurements, while the FWHM of J1905$+$1943 is constant throughout the campaign, the size of GRB\,221009A is clearly increasing in time.
Therefore, the variations observed for the GRB afterglow cannot be ascribed to a systematic effect due to an imprecise calibration.  Concerning the VLBA, no test was performed because of the lack of close enough check sources.

\section{Model of the projected size and proper motion of the forward and reverse shock}\label{apx:physical_model}

\subsection{Dynamics and size evolution}

In the following, we describe an approximate analytical model of the dynamics of the forward and reverse shocks, based on calculations similar to those of \citet{Kobayashi1999,Kobayashi2003,Yi2013}. The aim is to extend approaches such as those described in \citet{Oren2004} and \citet{Granot2005_GRB030329} by including the reverse shock, which was not considered there. We assume a cold external medium with a power law density profile $\rho=A m_\mathrm{p} (R/R_\star)^{-k}$, where $R$ is the radial distance from the progenitor, $m_\mathrm{p}$ is the proton mass and $A$ is the number density at a reference radius $R_\star = 5.5\times 10^{17}\,\mathrm{cm}$. With this definition, $A$ plays the role of either the homogeneous interstellar medium (ISM) number density, $A\equiv n$ if $k=0$, or that of the wind density parameter, $A\equiv A^\star$ if $k=2$.
We assume a simplified description of the jet as a cold, kinetic-energy-dominated shell with uniform initial bulk Lorentz factor $\Gamma_0 = 10^3\, \Gamma_{0,3}\gg 1$ and constant isotropic-equivalent kinetic luminosity $L=E/T$, where $E=10^{55}\,E_{55}\,\mathrm{erg}$ is the isotropic-equivalent jet energy and $T = T_{90}/(1+z) \approx 251\, T_{2.4}\,\mathrm{s}$ (where $T_{2.4}=T/(10^{2.4}\,\mathrm{s})$) is the lifetime of the central engine. The Sedov length associated with this shell is $\ell_\mathrm{S}=[(3-k)E/(4\pi A R_\star^k m_\mathrm{p} c^2)]^{1/(3-k)}$. As this shell expands into the external medium at relativistic speed, a FS arises, which sweeps the external medium moving with a Lorentz factor $\Gamma_\mathrm{FS,0}\sim \sqrt{2}\Gamma_0$. The shocked external medium resides in the region contained between the FS and the contact discontinuity (CD) that separates it from the jet material. Since this implies some deceleration of the jet material behind the CD as well, as soon as the ram pressure of such material overcomes the pressure in the jet (formally already at $R=0$ given our assumption of a cold jet), a RS also arises, which separates shocked from cold un-perturbed jet material. Let us indicate with numbers from 1 to 4 the un-perturbed external medium, shocked external medium, shocked jet and un-perturbed jet respectively, as usual. The RS is initially non-relativistic (i.e.\ the relative speed of regions 3 and 4 is $\beta_\mathrm{34}\ll 1$), but it can become relativistic before the RS crosses the whole jet if the condition \citep{Sari1995,Kobayashi1999,chevalier2000,Kobayashi2003,zou2005,Yi2013}
\begin{equation}
 \frac{E}{A}<\frac{4\pi m_\mathrm{p}c^2 R_\star^k}{3-k}\left(c T\right)^{3-k}\Gamma_0^{2(4-k)}\approx\left\lbrace\begin{array}{cr} 2.7\times 10^{60} \Gamma_{0,3}^8 T_{2.4}^3\,\mathrm{erg\,cm^3} & k=0 \\ 4.3\times 10^{58} \Gamma_{0,3}^3 T_{2.4}\,\mathrm{erg\,cm^3} & k=2\\
\end{array}\right. 
\label{eq:RRS_condition}
\end{equation}
is satisfied, in which case the jet deceleration is said to be in the `thick shell regime'. In the following we describe the dynamics in such regime, and we defer to later the treatment of the opposite, `thin shell' regime. For the homogeneous ISM case, $k=0$, the RS transitions from Newtonian to relativistic at a radius $R_\mathrm{N}\sim \ell_\mathrm{S}^{3/2}[(12cT)^{1/2}\Gamma_0^2]^{-1}\sim 4.2\times 10^{15}\,E_{55}^{1/2}A^{-1/2}\Gamma_{0,3}^{-2}T_{2.4}^{-1/2}\,\mathrm{cm}$, while in the wind case, $k=2$, the RS is always relativistic as long as condition \ref{eq:RRS_condition} holds.
As regions 2 and 3 decelerate due to the increasing amount of swept external medium mass, at some point the RS crosses the whole jet, at a radius
\begin{equation}
R_\otimes = (4(3-k)cT)^{1/(4-k)}\ell_\mathrm{S}^{(3-k)/(4-k)}.
\end{equation}
Before $R_\otimes$, regions 2 and 3 effectively expand at the same Lorentz factor $\Gamma$, whose evolution can be described approximately as
\begin{equation}
    \Gamma(R) \sim \left\lbrace\begin{array}{lr}\Gamma_0 & R\leq R_\mathrm{N} \\
\frac{\ell_\mathrm{S}^{(3-k)/4}}{\left[4(3-k)cT\right]^{1/4}}R^{-(2-k)/4} & R_\mathrm{N}<R\leq R_\otimes
\end{array}\right..
\end{equation}
The Lorentz factor of region 3 at the end of the RS crossing is therefore $\Gamma_\otimes = \left[\ell_\mathrm{S}/\left(4(3-k)cT\right)\right]^{(3-k)/[2(4-k)]}$. 
At radii larger than $R_\otimes$, the Lorentz factor of region 2 follows the \citet{Blandford1976} relativistic blastwave evolution, $\Gamma_2 \sim (R/\ell_\mathrm{S})^{-(3-k)/2}$. This holds as long as the lateral expansion of the shocked material in region 2 is negligible: numerical simulations and analytical arguments \citep[e.g.]{Kumar2003,vanEerten2010,DeColle2012,Lyutikov2012,Granot2012} show that such expansion has a very limited impact on the dynamics until region 2 becomes mildly relativistic, which justifies such assumption. In the homogeneous ISM case, $k=0$, the subsequent evolution of $\Gamma_3$ has been historically described phenomenologically \citep{Kobayashi2000} as $\Gamma_3 = \Gamma_\otimes (R/R_\otimes)^{-g}$, with $g$ being typically fixed at around $g\sim (7-2k)/2$ in the case of a non-relativistic RS, or at $g=(3-k)/2$ \citep{Kobayashi1999,Yi2013} in the case of a relativistic RS (when condition \ref{eq:RRS_condition} holds), based on insights from the numerical simulations described in \citet{Kobayashi1999} and \citet{Kobayashi2000}. Physically, the different evolution is likely related to the conversion of internal to kinetic energy in region 3 as it expands, which allows it to remain `attached' to region 2 as long as its temperature is relativistic. For historical reasons, in the case of a wind environment, $k=2$, the evolution in this phase has been always assumed to track that of the tail of the FS Blandford-McKee solution (i.e.\ $g=-3/2$), despite the lack of numerical simulations to compare to. We argue here that generally, as long as region 3 contains more internal than kinetic energy, its acceleration will keep it `attached' to region 2, and they will evolve following the \citet{Blandford1976} solution. As the internal energy conversion terminates, region 3 must eventually `detach' and expand backwards (as seen from the CD) into a rarefaction wave, and thus the evolution of $\Gamma_3$ with radius must steepen. 
In order to estimate the radius $R_\mathrm{det}$ at which regions 2 and 3 detach, we need to know the evolution of the internal energy in region 3, $E_\mathrm{int,3}$ (as measured in the comoving frame of region 3). From the first equation of thermodynamics, $d\ln E_\mathrm{int,3} = (1-\hat\gamma)d\ln V_3^\prime$, where $\hat\gamma$ is the adiabatic index and $V_3^\prime$ is the comoving volume of region 3. We assume $V_3^\prime \propto R^3/\Gamma_3$. Right after the shock crossing regions 2 and 3 still move together, hence we can assume $\Gamma_3\propto R^{-(3-k)/2}$, which leads to $E_\mathrm{int,3}=E_\mathrm{int,3,\otimes}\left(R/R_\otimes\right)^{(1-\hat\gamma)(9-k)/2}$. Taking the internal energy at the end of RS crossing to be $E_\mathrm{int,3,\otimes}\sim (\Gamma_{34,\otimes}-1)m_3 c^2 \sim (\Gamma_\otimes/\Gamma_0+\Gamma_0/\Gamma_\otimes)m_3 c^2/2$ (where $\Gamma_{34,\otimes}$ is the relative Lorentz factor of regions 3 and 4 at the RS crossing radius, and $m_3$ is the jet rest mass), we finally conclude that the effective dimensionless temperature in region 3 evolves as 
\begin{equation}
\Theta_3 \equiv E_\mathrm{int,3}/m_3 c^2 \sim \left[\frac{1}{2}\left(\frac{\Gamma_\otimes}{\Gamma_0}+\frac{\Gamma_0}{\Gamma_\otimes}\right)-1\right]\left(R/R_\otimes\right)^{(1-\hat\gamma)(9-k)/2}.    
\end{equation}
Assuming $\hat\gamma=4/3$ since the RS is relativistic, we finally obtain the detachment radius from the condition $\Theta_3(R_\mathrm{det})=1$, which yields
\begin{equation}
R_\mathrm{det}= \mathrm{max}\left\lbrace\left[\frac{1}{2}\left(\frac{\Gamma_\otimes}{\Gamma_0}+\frac{\Gamma_0}{\Gamma_\otimes}\right)-1\right]^{6/(9-k)} , 1 \right\rbrace R_\otimes,
\end{equation}
where the maximum function is introduced to account for cases where $\Theta_3<1$ at $R_\otimes$, in which case $R_\mathrm{det}=R_\otimes$. 

Based on these considerations, we model the evolution of $\Gamma_3$ after RS crossing as
\begin{equation}
    \Gamma_3(R)=\left\lbrace\begin{array}{lr}
         \Gamma_\otimes \left(\frac{R}{R_\otimes}\right)^{-(3-k)/2} & R_\otimes\leq R<R_\mathrm{det}  \\
         \Gamma_\otimes \left(\frac{R_\mathrm{det}}{R_\otimes}\right)^{-(3-k)/2}\left(\frac{R}{R_\mathrm{det}}\right)^{-(7-2k)/2} & R_\mathrm{det}\leq R\\ 
    \end{array}\right..
\end{equation}
The above relations completely specify the evolution of $\Gamma_2$ and $\Gamma_3$ with radius as a function of the Sedov length $\ell_\mathrm{S}$, initial Lorentz factor $\Gamma_0$ and jet duration $T$ for a given choice of $k$ in the thick shell regime. The thin shell regime is obtained \citep{Kobayashi1999} by setting all transition radii equal to the `deceleration' radius, $R_\mathrm{N}=R_\otimes=R_\mathrm{det}=\ell_\mathrm{S}/\Gamma_0^{2/(3-k)}$. The relation between the radius $R$ and the observer time for region $i\in \lbrace2,3\rbrace$ can be obtained by noting that most of the emission that the observer receives comes from material moving at an angle $\sim \Gamma_i^{-1}$ from the line of sight, for which the arrival time is $t_{\mathrm{obs}}/(1+z)=t(R)-R\beta_i(R)/c$, where $\beta_i=\sqrt{1-\Gamma_i^{-2}}$. The progenitor-frame time $t$ as a function of the radius can be obtained by integrating $t(R)=\int_0^R\mathrm{dR}/(\beta_i c)$. By numerically inverting the resulting relation between $t_{\mathrm{obs}}$ and $R$, we finally obtain $R_i(t_{\mathrm{obs}})$, and thus $\Gamma_2(t_{\mathrm{obs}})$ and $\Gamma_3(t_{\mathrm{obs}})$. The projected angular diameter of region $i=2,3$ is then approximately \citep[e.g.][]{Oren2004} 
\begin{equation}
s_{\mathrm m,i}(t_\mathrm{obs})\sim  2\frac{R_i(t_\mathrm{obs})}{d_\mathrm{A}}\times\left\lbrace\begin{array}{lr}
    \Gamma_i^{-1}(t_\mathrm{obs}) & \Gamma_i(t_\mathrm{obs})\geq 1/\theta_\mathrm{j} \\
    \theta_\mathrm{j} &  \Gamma_i(t_\mathrm{obs})<1/\theta_\mathrm{j} \\
\end{array}\right.,
\label{eq:model_source_diameter}
\end{equation}
where $\theta_\mathrm{j}$ is the jet half-opening angle. We find that the predicted size of the FS from this model matches that of the more refined model of \citet{Granot2005_GRB030329} within 10\% for $k\in \lbrace0,2\rbrace$ in the self-similar deceleration stage.

\subsection{Proper motion}\label{apx:proper_motion_model}

Unless the jet is observed perfectly on-axis, some apparent displacement of the projected image is expected. We discuss here the case $\theta_\mathrm{v}>\theta_\mathrm{j}$, which produces the largest such displacement.  As long as $\Gamma^{-1}<(\theta_\mathrm{v}-\theta_\mathrm{j})$, the observed emission is dominated by the border of the shock closest to the observer. Its apparent displacement $\Delta$ can be modelled effectively as that of a point source moving at $\sim c$ at an angle $\theta_\mathrm{v}-\theta_\mathrm{j}$ away from the line of sight, so that the displacement increases linearly in time, $\Delta \propto t_\mathrm{obs}$. For $(\theta_\mathrm{v}-\theta_\mathrm{j})<\Gamma^{-1}<(\theta_\mathrm{v}+\theta_\mathrm{j})$, the emission is dominated by material moving at $\sim 1/\Gamma$ from the line of sight, hence the displacement evolves as $\Delta(t_\mathrm{obs}) \sim R(t_\mathrm{obs})/\Gamma(t_\mathrm{obs})d_\mathrm{A}$, with $R(t_\mathrm{obs})$ being the same as in the on-axis case. Eventually, for $\Gamma^{-1}>(\theta_\mathrm{v}+\theta_\mathrm{j})$, the emission is dominated by the material at the shock border farthest from the observer, and the displacement is therefore $\Delta(t_\mathrm{obs}) \sim R(t_\mathrm{obs})\sin(\theta_\mathrm{v}+\theta_\mathrm{j})/d_\mathrm{A}$. The transition times $t_\mathrm{j,obs,\pm}$ that separate the three regimes described above can be obtained by setting $\Gamma(t_\mathrm{obs})=(\theta_\mathrm{v}\pm\theta_\mathrm{j})^{-1}$ in the on-axis case, which we do numerically. 

The model described here neglects the effects of lateral expansion of the shock and of a non-trivial jet structure outside the `core' of half-opening angle $\theta_\mathrm{j}$. The former would generally slow down the evolution, so that the displacement predicted by this model can be considered as an upper limit. The latter would change (generally steepen) the slope of the displacement with time before the time $t_\mathrm{j,obs,-}$ at which the jet core starts coming into sight, but not thereafter. For the most likely parameters, our observations are at $t_\mathrm{obs}>t_\mathrm{j,obs,-}$, so that the effects of a jet structure are unimportant for this particular source.

\begin{figure*}
    \centering
    \includegraphics[width=\textwidth]{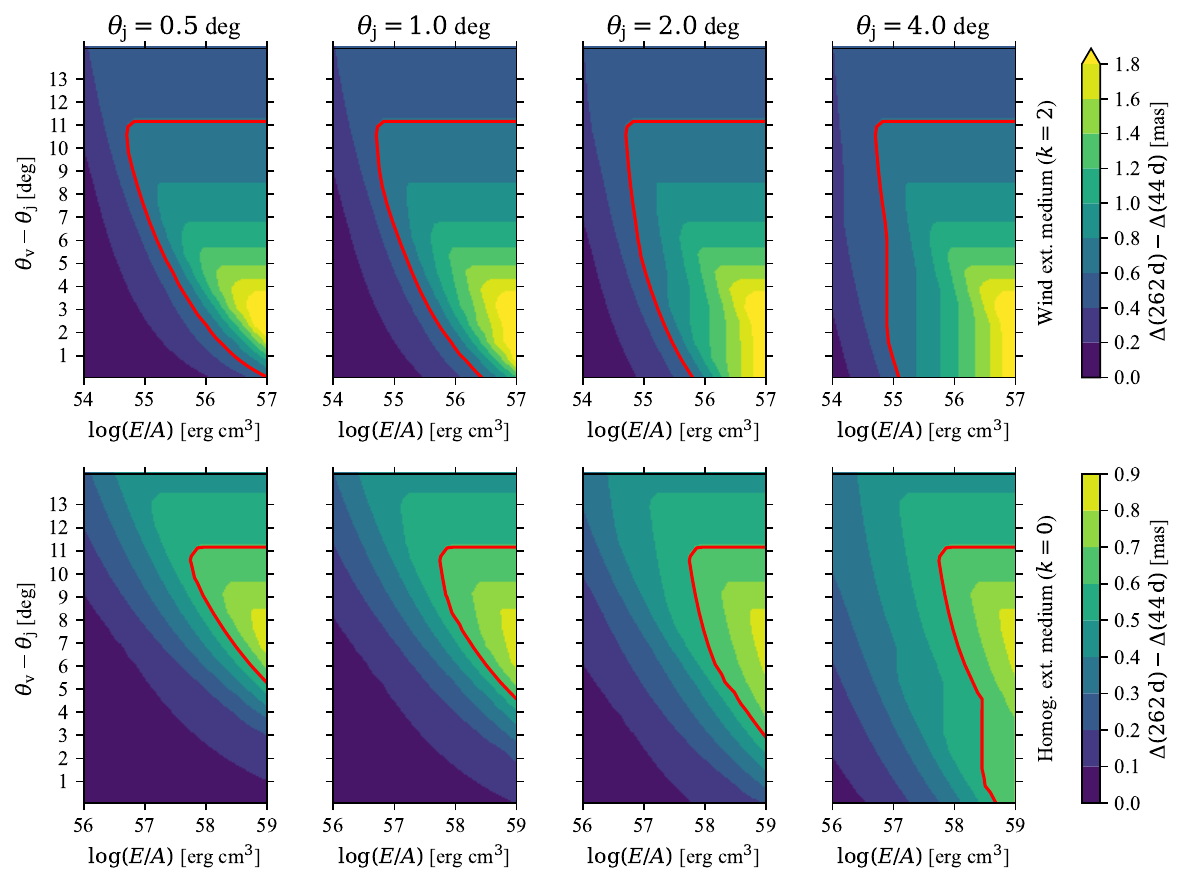}
    \caption{Constraint on the viewing angle from the absence of an observed source apparent displacement in our VLBA observations. In each panel, filled contours show the displacement of the centre of the fitted Gaussian expected between 44 d and 262 d, color coded as shown in the colorbar on the right, as a function of the $E/A$ ratio and of the off-edge viewing angle $\theta_\mathrm{v}-\theta_\mathrm{j}$. The red contour shows $\Delta(262\,\mathrm{d})-\Delta(44\,\mathrm{d})=0.6\,\mathrm{mas}$, which represents the largest displacement compatible at 1 $\sigma$ with our observations. The red contour hence contains the excluded parameter region. The upper panel row refers to a wind-like external medium, while the lower row refers to a homogeneous external medium. Each column assumes a different jet half-opening angle, given at the top of the column.}
    \label{fig:viewing_angle_constraint}
\end{figure*}

Figure \ref{fig:viewing_angle_constraint} shows the displacement predicted by such a model between 44 d and 262 d, assuming the emission to be dominated by the FS (which produces the largest displacement, and dominates the VLBA data according to our interpretation) for different assumptions on $\theta_\mathrm{j}$ (varying across columns) and on the external medium power law index (top row: $k=2$; bottom row: $k=0$), as a function of the off-edge viewing angle $\theta_\mathrm{v}-\theta_\mathrm{j}$ and of the energy to density ratio $E/A$. These predictions show that our upper limit on the observed displacement only excludes off-edge viewing angles between a few degrees and around 11 degrees, combined with large energy to density ratios $E/A\gtrsim 10^{55}\,\mathrm{erg\,cm^3}$ for $k=2$, or rather extreme $E/A\gtrsim 10^{58}\,\mathrm{erg\,cm^3}$ for $k=0$. Viewing angles larger than $\theta_\mathrm{v}-\theta_\mathrm{j}\sim 11\,\mathrm{deg}$ cannot be constrained because in that case the shock is in the `point-source at the jet edge' regime all the way to $262$ d, with a rather small apparent transverse velocity. On the other hand, such a large viewing angle would be very unlikely given the huge $\upgamma$-ray isotropic-equivalent energy of this source. 
\end{appendix}

\end{document}